\input harvmac

%%%%%%%%%%%%%%%%%%

%
%  This inputs the macro package epsf.tex
%
\ifx\epsfbox\UnDeFiNeD\message{(NO epsf.tex, FIGURES WILL BE IGNORED)}
\def\figin#1{\vskip2in}% blank space instead
\else\message{(FIGURES WILL BE INCLUDED)}\def\figin#1{#1}\fi
\def\ifig#1#2#3{\xdef#1{fig.~\the\figno}
\goodbreak\midinsert\figin{\centerline{#3}}%
\smallskip\centerline{\vbox{\baselineskip12pt
\advance\hsize by -1truein\noindent\footnotefont{\bf Fig.~\the\figno:}
#2}}
\bigskip\endinsert\global\advance\figno by1}

\def\ifigure#1#2#3#4{
\midinsert
\vbox to #4truein{\ifx\figflag\figI
\vfil\centerline{\epsfysize=#4truein\epsfbox{#3}}\fi}
\narrower\narrower\noindent{\footnotefont
{\bf #1:}  #2\par}
\endinsert
}

%**************************************************

\def\IC{{\ \hbox{{\rm I}\kern-.6em\hbox{\bf C}}}}
\def\IR{{\hbox{{\rm I}\kern-.2em\hbox{\rm R}}}}
\def\IZ{{\hbox{{\rm Z}\kern-.4em\hbox{\rm Z}}}}

\def\sIR{{\hbox{{\sevenrm I}\kern-.2em\hbox{\sevenrm R}}}}

\def\p{p_{11}}

\Title{RU-97-76}{\vbox{\centerline{Matrix Theory
}}}

\centerline{\it T. Banks~$^1$}
\medskip
\centerline{$^1$Department of Physics and Astronomy}
\centerline{Rutgers University, Piscataway, NJ 08855-0849}
\centerline{\tt banks@physics.rutgers.edu}

\bigskip

\medskip

\noindent
This is an expanded version of talks given by the author at the Trieste
Spring School on Supergravity and Superstrings in April of 1997 and at
the accompanying workshop.  The manuscript is intended to be a
mini-review of Matrix Theory.   The motivations and some of the evidence
for the theory are presented, as well as a clear statement of the
current puzzles about compactification to low dimensions.

\hyphenation{Min-kow-ski}
\Date{September 1997}

\def\symd{SYM_{d+1}}
\def\lp{l_{11}}
\def\lqq{\lq\lq}
\def\rqq{\rq\rq}
\def\nto{N\rightarrow \infty}
\def\td{{\bf T^d}}
\def\ttwo{{\bf T^2}}
\def\tdd{{\bf \tilde{T^d}}}
\def\tr{\hbox{Tr}}
\def\p{p^+}
\newsec{\bf INTRODUCTION}
\subsec{M Theory}

M theory is a misnomer.  It is not a theory, but rather a collection of
facts and arguments which suggest the existence of a theory.   The
literature on the subject is even somewhat schizophrenic about the
precise meaning of the term M theory.  For some authors it represents
another element in a long list of classical vacuum configurations of
\lq\lq the theory formerly known as String \rq\rq .  For others it is
the overarching {\it ur} theory itself.  We will see that this dichotomy
originates in a deep question about the nature of the theory, which we
will discuss extensively, but not resolve definitively.  In these
lectures we will use the term M theory to describe the theory which
underlies the various string perturbation expansions. We will
characterize the eleven dimensional quantum theory whose low energy
limit is supergravity (SUGRA) with phrases like \lq\lq the eleven
dimensional limit of M theory \rq\rq .

M theory arose from a collection of arguments indicating that the
strongly coupled limit of Type IIA superstring theory is described at
low energies by eleven dimensional supergravity \ref\mth{
M.J. Duff, P. Howe,
T. Inami, K.S. Stelle, "Superstrings in D=10 
{}from supermembranes in D=11", Phys. Lett. B191 (1987) 70\semi
M.J. Duff, J. X. Lu, "Duality rotations in membrane theory", Nucl. 
Phys. B347 (1990) 394 \semi M.J. Duff, R. Minasian, James T. Liu, "Eleven-dimensional origin of 
string/string duality: a one-loop test", Nucl. Phys. B452 (1995) 261 \semi
C. Hull and P. K. Townsend, "Unity of superstring dualities",
Nucl. Phys. { B438} (1995) 109,
hep-th/9410167
\semi
P. K. Townsend, ``The Eleven-Dimensional Supermembrane Revisited''
 Phys. Lett. B350 (1995) 184 hep-th/9501068 \semi
"String-membrane duality in seven dimensions", Phys. Lett. { 354B} 
(1995) 247, hep-th/9504095 \semi
C. Hull and P. K. Townsend,
``Enhanced gauge symmetries in superstring theories", Nucl. Phys. { B451} 
(1995) 525, hep-th/9505073\semi
E. Witten ``String Theory Dynamics in Various Dimensions",
Nucl.Phys. B443 (1995) 85 hep-th/9503124.
} .  Briefly,
and somewhat anachronistically, the argument hinges on the existence of
D0 brane solitons of Type IIA string theory \ref\dbranes{J.Polchinski, 
{\it TASI Lectures on D-Branes}, hep-th/9611050 .}.  These are
pointlike (in the ten dimensional sense) , Bogolmonyi-Prasad-Sommerfield
(BPS) states \foot{For a review of BPS states and extensive references,
see the lectures of J. Louis in these proceedings.}, with mass ${1\over
l_S g_S}$.  If one makes the
natural assumption\ref\witten{E. Witten, Nucl.Phys. B443 (1995) 85
hep-th/9503124 .} that there is a threshold bound state of $N$
D0 branes for any $N$, then one finds in the strong coupling limit a
spectrum of low energy states coinciding with the spectrum of eleven
dimensional supergravity\foot{The authors of \ref\sethi{S. Sethi,
M. Stern, hep-th/9705046; M.Porrati, A. Rozenberg, hep-th/9708119 .} 
have recently proven the 
existence of the threshold bound state for $N=2$, and $N$ prime 
respectively.} .  The general properties of M theory are
derived simply by exploiting this fact, together with the assumed
existence of membranes and fivebranes of the eleven dimensional
theory\foot{It is often stated that the fivebrane is a smooth soliton in 
11 dimensional SUGRA
and therefore its existence follows from the original hypothesis. However, 
the scale of variation of the soliton fields is $\lp$, the scale at which
the SUGRA approximation breaks down, so this argument should be taken with
a grain of salt.}, on
various partially compactified eleven manifolds \ref\mthexamples{
J.H. Schwarz, Phys.Lett. B367 (1996) 97, hep-th/9510086; K.Dasgupta,
S.Mukhi, Nucl.Phys. B465 (1996) 399, hep-th/9512196; E.Witten
Nucl.Phys. B463 (1996) 383, hep-th/9512219; A.Sen Phys.Rev. D53 (1996)
6725, hep-th/9602010; K.Becker, M.Becker Nucl.Phys. B472 (1996) 221,
hep-th/9602071; S.Ferrara, R.R.Khuri, \break R.Minasian, Phys.Lett. B375 (1996)
81, hep-th/9602102; A.Kumar, K.Ray, Phys.Rev. D54 (1996) 1647,
hep-th/9602144; O.Aharony, J.Sonnenschein, S.Yankielowicz,
Nucl.\break Phys. B474 (1996) 309, hep-th/9603009; P.Aspinwall,
Nucl.Phys.Proc.Suppl. 46 (1996) 30, hep-th/9508154; P.Horava, E.Witten,
Nucl.Phys. B475 (1996) 94, hep-th/9603142, Nucl. Phys. B460,(1996), 506,
hep-th/9510209; A.Sen, Mod.Phys.Lett. A11
(1996) 1339, hep-th/9603113; B.S.Acharya, hep-th/96041-33; K.Becker,
M.Becker, Nucl.Phys. B477 (1996) 155, hep-th/9605053; R.Gopakumar,
S.Mukhi, Nucl.Phys. B479 (1996) 260, hep-th/9607057; J.Schwarz,
Nucl.Phys.Proc.\break Suppl. 55B (1997) 1, hep-th/9607201 ({\it and references
therein }); M.J.Duff, Int.J.Mod.\break Phys. A11 (1996), 5623, hep-th/9608117;
P.K.Townsend, Talks given at Summer School in High Energy Physics and
Cosmology, Trieste, June 1996, hep-th/9612121; P.Aspinwall,
hep-th/9707014; S.Roy, Nucl. Phys. B498 (1997), 175.}.

At this point we can already see the origins of the dichotomic attitude
to M theory which can be found in the literature.  In local field
theory, the behavior of a system on a compact space is essentially
implicit in its infinite volume limit.  Apart from well understood
topological questions which arise in gauge theories, 
the degrees of freedom in the compactified theory
are a restriction of those in the flat space limit.  From this point of
view it is natural to think of the eleven dimensional limiting theory as
the underlying system from which all the rest of string theory is to be
derived.  The evidence presented for M theory in \mthexamples\ can be
viewed as support for this point of view.   

On the other hand, it is important to realize that the contention that all the
degrees of freedom are implicit in the infinite volume theory is far
from obvious in a theory of extended objects.  Winding and wrapping
modes of branes of various dimensions go off to infinite energy as the
volume on which they are wrapped gets large.  If these are fundamental
degrees of freedom, rather than composite states built from local
degrees of freedom, then the prescription for compactification involves
the addition of new variables to the Lagrangian.  It is then much less
obvious that the decompactified limit is the {\it ur} theory from which
all else is derived.   It might be better to view it as \lq\lq just
another point on the boundary of moduli space \rq\rq .

\subsec{M is for Matrix Model}

The purpose of these lecture notes is to convince the reader that {\it
Matrix Theory} is in fact the theory which underlies the various string
perturbation expansions which are currently known.  We will also argue
that it has a limit which describes eleven dimensional Super-Poincare
invariant physics (which is consequently equivalent to SUGRA at low
energies).  The theory is still in a preliminary stage of development,
and one of the biggest lacunae in its current formulation is precisely 
the question raised about M theory in the previous paragraph.  We do not
yet have a general prescription for compactification of the theory and
are consequently unsure of the complete set of degrees of freedom which
it contains.  In Matrix Theory this question has a new twist, for the
theory is defined by a limiting procedure in which the number of degrees
of freedom is taken to infinity.  It becomes somewhat difficult to
decide whether the limiting set of degrees of freedom of the
compactified theory are a subset of those of the uncompactified theory.
 Nonetheless, for a variety of compactifications, Matrix Theory provides a
nonperturbative definition of string theory which incorporates much of
string duality in an explicit Lagrangian formalism and seems to
reproduce the correct string perturbation expansions of several
different string theories in different limiting situations.

We will spend the bulk of this review trying to explain what is right
about Matrix Theory.  It is probably worth while beginning with a list
of the things which are {\it wrong} with it.

\item{1.} First and foremost, Matrix Theory is formulated in the light
cone frame.  It is constructed by building an infinite momentum frame (IMF)
boosted along a compact direction by starting from a frame with $N$ units of
compactified momentum and taking $N$ to infinity.  
Full Lorentz invariance is not obvious and will arise, if
at all, only in the large $N$ limit.  It also follows from this that
Matrix Theory is {\it not background independent}.  Our matrix
Lagrangians will contain parameters which most string theorists believe
to be properly viewed as expectation values of dynamical fields. 
In IMF dynamics, such zero
 momentum modes have infinite frequency and are frozen into a fixed 
configuration.
In a semiclassical expansion, 
quantum corrections to the potential which determines the allowed background
configurations show up as divergences at zero longitudinal momentum.  We
will be using a formalism in which these divergences are related to the 
large $N$ 
divergences in a matrix Hamiltonian.

\item{2.}A complete prescription for enumeration of allowed backgrounds
has not yet been found.  At the moment we have only a prescription for
toroidal compactification of Type II strings on tori of dimension $\leq 4$ 
and the beginning of a
prescription for toroidal compactification of heterotic strings on tori of 
dimension $\leq 3$ (this
situation appears to be changing as I write).

\item{3.}Many of the remarkable properties of Matrix Theory appear to be
closely connected to the ideas of Noncommutative Geometry
\ref\noncom{A. Connes, {\it Noncommutative Geometry}, Academic Press 1994.}
. These connections have so far proved elusive.  
\item{4.}Possibly related to the previous problem is a serious esthetic
defect of Matrix Theory.  String theorists have long fantasized about a
beautiful new physical principle which will replace Einstein's marriage
of Riemannian geometry and gravitation.  Matrix theory most emphatically
does not provide us with such a principle.  Gravity and geometry emerge
in a rather awkward fashion, if at all.  Surely this is the major defect
of the current formulation, and we need to make a further conceptual
step in order to overcome it.

In the sections which follow, we will take up the description of Matrix
theory from the beginning.  We first describe the general ideas of
holographic theories in the infinite momentum frame (IMF), and argue
that when combined with maximal supersymmetry they lead one to a unique 
Lagrangian for the fundamental degrees of freedom (DOF) in flat, infinite, 
eleven dimensional spacetime.  We then show
that the quantum theory based on 
this Lagrangian contains the Fock space of eleven dimensional
supergravity (SUGRA), as well as metastable states representing large
semiclassical supermembranes.  
Section III describes the prescription for compactifying this eleven
dimensional theory on tori and discusses the extent to which the DOF of
the compactified theory can be viewed as a subset of those of the eleven
dimensional theory.  Section IV shows how to extract Type IIA and IIB
perturbative string theory from the matrix model Lagrangian and
discusses T duality and the problems of compactifying many dimensions.

Section V contains the matrix model description of Horava-Witten domain
walls and $E_8 \times E_8$ heterotic strings.   Section VI is devoted to
BPS p-brane solutions to the matrix model.   Finally, in the conclusions,
we briefly list some of the important topics not covered in this 
review\foot{We note here
that a major omission will be the important but as yet incomplete 
literature on Matrix
Theory on curved background spaces.  A fairly comprehensive set of
references can be found
in \ref\mrd{M.R.Douglas, Talk given at Strings 97, Amsterdam, June 1997}
and citations therein.} ,
and suggest directions for further research.

\newsec{\bf HOLOGRAPHIC THEORIES IN THE IMF}

\subsec{General Holography}

For many years, Charles Thorn \ref\thorn{ 
 C.B. Thorn, Proceedings of
Sakharov Conf. 
on Physics Moscow (1991) 447-454, hep-th/9405069 .} has championed an
approach to
nonperturbative string theory based on the idea of {\it string bits}.
Light cone gauge string theory can be viewed as a parton model in an IMF
along a compactified spacelike dimension, whose partons, or fundamental
degrees of freedom carry only the lowest allowed value of longitudinal
momentum. In perturbative string theory, 
this property, which contrasts dramatically with the properties of partons
in local field theory, follows from the fact that longitudinal momentum
is (up to an overall factor) the {\it length} of a string in the IMF.
Discretization of the longitudinal momentum is thus equivalent to a world
sheet cutoff in string theory and the partons are just the smallest bits
of string.  Degrees of freedom with larger longitudinal momenta are
viewed as composite objects made out of these fundamental bits.  Thorn's
proposal was that this property of perturbative string theory should be
the basis for a nonperturbative formulation of the theory.

Susskind \ref\lenholo{ L. Susskind,  J. Math. Phys.  36
1995 6377, hep-th/940989.} realized that this property of string theory
suggested that string theory obeyed the {\it holographic principle},
which had been proposed by `t Hooft \ref\thooft{
 G. 't
Hooft, {\it Dimensional Reduction in Quantum Gravity}, Utrecht preprint
THU-93/26, gr-qc/9310026 } as the basis of a
quantum theory of black holes.  The `t Hooft-Susskind holographic
principle states that the fundamental degrees of freedom of a consistent
quantum theory including gravity must live on a $d-2$ dimensional
transverse slice of $d$ dimensional space-time.  This is equivalent to
demanding that they carry only the lowest value of longitudinal momentum, so that wave
functions of composite states are described in terms of purely transverse
parton coordinates.
`t Hooft and Susskind
further insist that the DOF obey the Bekenstein \ref\bek{
J. D. Bekenstein, Phys.Rev. D 49 (1994), 6606} bound: the
transverse density of DOF should not exceed one per Planck area.
Susskind noted that this bound was not satisfied by the wave functions
of perturbative string theory, but that nonperturbative effects became
important before the Bekenstein bound was exceeded.  He conjectured that
the correct nonperturbative wave functions would exactly saturate the
bound.  We will see evidence for this conjecture below.
It seems clear that this part of the holographic principle may be a
dynamical consequence of Matrix Theory but is not one of its underlying
axioms. 

In the IMF, the full holographic principle leads to an apparent paradox.
As we will review in a moment, the objects of study in IMF physics are
composite states carrying a finite fraction of the total longitudinal
momentum.  The holographic principle requires such states to contain an
infinite number of partons.   The Bekenstein bound requires these
partons to take up an area in the transverse dimensions which grows like
$N$, the number of partons.  

On the other hand, we are trying to
construct a Lorentz invariant theory which reduces to local field theory
in typical low energy situations.  
Consider the scattering of two objects at low center of mass energy and
large impact parameter in their center of mass frame in flat spacetime.
This process must be described by local field theory to a very good
approximation.   Scattering amplitudes must go to zero in this low
energy, large impact parameter regime.
In a Lorentz invariant holographic 
theory, the IMF wave functions of the two objects have infinite extent
in the transverse dimensions.  Their wave functions overlap.   Yet
somehow the parton clouds do not interact very strongly even when they
overlap.  We will see evidence that the key to resolving this paradox is
supersymmetry (SUSY), and that SUSY is the basic guarantor of
approximate locality at low energy.   

\subsec{Supersymmetric Holography}

In any formulation of a Super Poincare invariant\foot{It is worth
spending a moment to explain why one puts so much emphasis on Poincare
invariance, as opposed to general covariance or some more sophisticated
curved spacetime symmetry.  The honest answer is that this is what we
have at the moment.  Deeper answers might have to do with the 
holographic principle, or with noncommutative geometry.  
In a holographic theory in asymptotically flat spacetime, one can always
imagine choosing the transverse slice on which the DOF lie to be in the
asymptotically flat region, so that their Lagrangian should be
Poincare invariant.  Another approach to understanding how curved spacetime
could arise comes from noncommutative geometry. 
The matrix model approach to noncommutative
geometry utilizes coordinates which live in a linear space of matrices. 
Curved spaces arise by integrating out some of these linear variables.}
 quantum theory which is
tied to a particular class of reference frames, some of the generators
of the symmetry algebra are {\it easy} to write down, while others are
{\it hard}.  Apart from the Hamiltonian which defines the quantum
theory, the easy generators are those which preserve the equal time
quantization surfaces.
We will try to construct a holographic IMF theory by taking
the limit of a theory with a finite number of DOF.  As a consequence,
longitudinal boosts will be among the {\it hard} symmetry
transformations to implement, along with the null-plane rotating Lorentz
transformations which are the usual bane of IMF physics.
These should only become manifest in the $N \rightarrow \infty$ limit.
The easy generators form the Super-Galilean algebra.  It consists of
transverse rotations $J^{ij}$, transverse boosts, $K^i$ and
supergenerators. Apart from the obvious rotational commutators, the
Super-Galilean algebra has the form:
\eqn\sga{   
\eqalign{[Q_{\alpha},Q_{\beta}]_+ &= \delta_{\alpha\beta} H \cr
         [q_A,q_B]_+ &= \delta_{AB}P_L \cr
         [Q_{\alpha} ,q_A ] &= \gamma^i_{A\alpha}  P_i \cr}}
  
\eqn\sgb{[K^i , P^j ] = \delta^{ij} P^+}

We will call the first and second lines of \sga\ the dynamical and
kinematical parts of the
supertranslation algebra respectively.  Note that we work in $9$
transverse dimensions, as is appropriate for a theory with eleven
spacetime dimensions.  The tenth spatial direction is the longitudinal
direction of the IMF.  We imagine it to be compact, with radius $R$. The
total longitudinal momentum is denoted $N/R$.
The Hamiltonian is the generator of translations
in light cone time, which is the difference between the IMF energy and
the longitudinal momentum.  

The essential simplification of the IMF follows from thinking about the
dispersion relation for particles
\eqn\disprel{E = \sqrt{P_L^2 + {\bf P_{\perp}}^2 + M^2} \rightarrow
\vert P_L \vert + {{{\bf P_{\perp}}^2 + M^2} \over 2 P_L}
}
The second form of this equation is {\it exact} in the IMF.  It shows us
that particle states with negative or vanishing longitudinal momentum
are eigenstates of the IMF Hamiltonian, $E - P_L$  with infinite 
eigenvalues.  Using
standard renormalization group ideas, we should be able to integrate
them out, leaving behind a local in time, Hamiltonian, formulation of
the dynamics of those degrees of freedom with positive longitudinal
momenta.  In particular, those states which carry a finite fraction of
the total longitudinal momentum $k/R$ with $k/N$ finite as $\nto$, will
have energies which scale like $1/N$.  It is these states which we expect
to have Lorentz invariant kinematics and dynamics in the $\nto$ limit.
In a holographic theory, they will be composites of fundamental partons with
longitudinal momentum $1/R$.

The dynamical SUSY algebra \sgb\ is very difficult to satisfy.  Indeed
the known representations of it are all theories of free particles.  To
obtain interacting theories one must generalize the algebra to
\eqn\gaugealg{\{ Q_{\alpha} , Q_{\beta} \} = \delta_{\alpha\beta} + Y^A G^A}
where $G^A$ are generators of a gauge algebra, which annihilate physical
states.    The authors of \ref\susyqm{
M.Claudson, M.B.Halpern, Nucl. Phys. B250, (1985), 689;
M. Baake, P. Reinicke, V. Rittenberg, J. Mathm. Phys., 26,
(1985), 1070; R. Flume, Ann. Phys. 164, (1985), 189.} have shown that if 

\item{1.} The DOF transform in the adjoint representation of the gauge
group.

\item{2.} The SUSY generators are linear in the canonical momenta of
both Bose and Fermi variables.

\item{3.} There are no terms linear in the bosonic momenta in the
Hamiltonian.

\noindent
then the unique representation of this algebra with a finite number of
DOF is given by the dimensional reduction of $9 + 1$ dimensional SUSY
Yang Mills ($SYM_{9 + 1}$) to $0 + 1$ dimensions.  The third hypothesis
can be eliminated by using the restrictions imposed by the rest of the
super Galilean algebra.   These systems in fact possess the full
Super-Galilean symmetry, with kinematical SUSY generators given by 
\eqn\kinsusy{q_{\alpha} = Tr\  \Theta_{\alpha},} 
where $\Theta_{\alpha}$ are the fermionic superpartners of the gauge field.
Indeed, I believe that the unique interacting Hamiltonian with the full 
super Galilean symmetry in $9$ transverse dimensions is given by the
dimensionally reduced SYM theory.  Note in particular that any sort of
naive nonabelian generalization of the Born-Infeld action would violate
Galilean boost invariance, which is an exact symmetry in the IMF\foot{It
is harder to rule out Born-Infeld type corrections with coefficients
which vanish in the large $N$ limit.}.
Any corrections to the SYM Hamiltonian must vanish for Abelian 
configurations of the variables.   The restriction to variables
transforming in the adjoint representation can probably be removed as
well.  We will see below that fundamental representation fields can
appear in Matrix theory, but only in situations with less than maximal
SUSY.

In order to obtain an interacting Lagrangian in which the number of
degrees of freedom can be arbitrarily large, we must restrict attention
to the classical groups $U(N)$, $O(N)$, $USp(2N)$.  For reasons which are
not entirely clear, the only sequence which is realized is $U(N)$.
The orthogonal and symplectic groups do appear, but again only in
situations with reduced SUSY.

More work is needed to sharpen and simplify these theorems about
possible realizations of the maximal Super Galilean algebra.  It is
remarkable that the holographic principle and supersymmetry are so
restrictive and it behooves us to understand these restrictions better
than we do at present.  However, if we accept them at face value, these
restrictions tell us that an interacting, holographic eleven dimensional SUSY theory
, with a finite number of degrees of freedom, is essentially unique.

To understand this system better, we now present an alternative
derivation of it, starting from weakly coupled Type IIA string theory.
The work of Duff, Hull and Townsend, and Witten \mth , established the
existence of an eleven dimensional quantum theory called M theory.
Witten's argument proceeds 
by examining states which are charged under the Ramond-Ramond one form
gauge symmetry.  The fundamental charged object is a $D0$ brane
\dbranes\ , whose mass is $1/g_S l_S$.  $D0$ branes are BPS states.
If one hypothesizes the existence of 
a threshold bound state of $N$ of these particles\foot{For $N$ prime,
this is not an hypothesis, but a theorem, proven  in \sethi\ .},
and takes into account the degeneracies implied by SUSY, one finds a
spectrum of states exactly equivalent to that of eleven dimensional
SUGRA compactified on a circle of radius $R = g_S l_S$.  

The low energy effective Lagrangian of Type IIA string theory is in fact
the dimensional reduction of that of $SUGRA_{10 + 1}$ with the string
scale related to the eleven dimensional Planck scale by $l_{11} =
g_S^{1/3} l_S$.  These relations are compatible with a picture of the
IIA string as a BPS membrane of SUGRA, with tension $\sim l_{11}^{-3}$
wrapped around a circle of radius $R$.  

In \ref\bfss{T.Banks, W.Fischler, S.Shenker, L.Susskind, Phys.Rev. D55
(1997) 112, \break hep-th/9610043. }\ it was pointed out that the identification of the strongly
coupled IIA theory with an eleven dimensional theory showed that the 
holographic philosophy was applicable to this highly nonperturbative 
limit of string theory.  Indeed, if IIA/M theory duality is correct, the
momentum in the tenth spatial dimension is identified with Ramond-Ramond
charge, and is carried only by $D0$ branes and their bound states.
Furthermore, if we take the $D0$ branes to be the fundamental
constituents, then they carry only the lowest unit of longitudinal
momentum.   In an ordinary reference frame, one also has anti-$D0$
branes, but in the IMF the only low energy DOF will be positively
charged $D0$ branes\foot{A massless particle state with any nonzero
transverse momentum will eventually have positive longitudinal momentum if
it is boosted sufficiently.  Massless particles with exactly zero transverse
momentum are assumed to form a set of measure zero.  If all transverse
dimensions are compactified this is no longer true, and such states may have 
a role to play.}.  

In this way of thinking about the system, one goes to the IMF by adding
$N$ $D0$ branes to the system and taking $\nto$.  The principles of IMF 
physics seem to 
tell us that a
complete Hamiltonian for states of finite light cone energy 
can be constructed using only $D0$  branes as DOF.
This is not quite correct.

In an attempt to address the question of the existence of threshold
bound states of $D0$ branes, Witten\ref\edbound{E.Witten,
Nucl.Phys. B460 (1996) 335, hep-th/9510135.} constructed a
Hamiltonian for low energy processes involving zero branes at relative
distances much smaller than the string scale in weakly coupled string
theory.   The Hamiltonian and SUSY generators have the form
\eqn\kinsusy{q_{\alpha} = \sqrt{R^{-1}} \tr \Theta}
\eqn\dynsusy{Q_{\alpha} = \sqrt{R} \tr [ \gamma^i_{\alpha\beta} P^i + i
[X^i ,X^j ] \gamma^{ij}_{\alpha\beta}] \Theta_{\beta}}
\eqn\ham{H=R\ \tr \left\{{{{ \Pi_i \Pi_i}\over 2}
- {1\over 4}\ [X^i,X^j]^2}+{\theta^T} \gamma_i [\Theta,X^i]
\right\}}
where we have used the scaling arguments of \ref\kpdfs{U.Danielsson,
G.Ferretti, B.Sundborg, Int.J.Mod.Phys. A11, (1996), 5463; hep-th/9603081;
D.Kabat, P.Pouliot, Phys.Rev.Lett. 77 (1996) 1004, hep-th/9603127.} to
eliminate the string coupling and string scale in favor of the eleven
dimensional Planck scale.  We have used conventions in which the
transverse coordinates $X^i$ have dimensions of length, and $\lp = 1$.
The authors of \ref\dkps{M.R.Douglas,
D.Kabat, P.Pouliot, S.H.Shenker, Nucl.Phys. B485 (1997) 85,
hep-th/9608024.}
 showed that this
Hamiltonian remained valid as long as the transverse velocities of the
zero branes remained small.  We emphasize that this was originally
interpreted as an ordinary Hamiltonian for a few zero branes in an
ordinary reference frame.  As such, it was expected to have relativistic
corrections, retardation corrections {\it etc.}.  However, when we go to
the IMF by taking $\nto$ we expect the velocities of the zero branes to
go to zero parametrically with $N$ (we will verify this by a dynamical
calculation below).  Furthermore, SUSY forbids any renormalization of
the terms quadratic in zero brane velocity,  Thus, it is plausible to
conjecture that this is the exact Hamiltonian for the zero brane system
in the IMF, independently of the string coupling.

The uninitiated (surely there are none such among our readers) may be
asking where the zero branes are in the above Hamiltonian.  The bizarre
answer is the following:  {\it The zero brane transverse coordinates,
and their superpartners, are the diagonal matrix elements of the Hermitian
matrices $X^i$ and $\Theta$.  The off diagonal matrix elements are
creation and annihilation operators for
the lowest lying states of open strings stretched between zero branes.}
The reason we cannot neglect the open string states is that the system
has a $U(N)$ gauge invariance (under which the matrices transform in the
adjoint representation), which transforms the zero brane coordinates
into stretched open strings and vice versa.  It is only when this
invariance is \lq\lq spontaneously broken \rq\rq by making large
separations between zero branes, that we can disentangle the diagonal
and off diagonal matrix elements.  A fancy way of saying this (which we
will make more precise later on, but whose full implications have not
yet been realized) is to say that the matrices $X^i$ and $\Theta$ are
the supercoordinates of the zero branes in a {\it noncommutative
geometry}. 

To summarize: general IMF ideas, coupled with SUSY nonrenormalization
theorems, suggest that the exact IMF 
Hamiltonian of strongly coupled Type IIA string theory is given by the
large $N$ limit of the Hamiltonian \ham\ .  The longitudinal momentum is
identified with $N/R$ and the SUSY generators are given by \kinsusy\ and
\dynsusy\ .  This is precisely the Hamiltonian which we suggested on
general grounds above.

\subsec{Exhibit A}

We do not expect the reader to come away convinced by the arguments
above\foot{Recently, Seiberg\ref\natiproof{N.Seiberg, hep-th/9710009.} has come up with a proof
that Matrix Theory is indeed the exact Discrete Light Cone Quantization
of M theory.}.  The rest of this review will be a presentation of the evidence
for the conjecture that the matrix model Hamiltonian \ham\ indeed
describes a covariant eleven dimensional quantum mechanics with all the
properties ascribed to the mythical M theory, and that various
compactified and orbifolded versions of it reduce in appropriate limits
to the weakly coupled string theories we know and love.  This subsection
will concentrate on properties of the eleven dimensional theory.

First of all we show that the $\nto$ limit of the matrix model contains
the Fock space of eleven dimensional SUGRA.  The existence of single
supergraviton states follows immediately from the hypothesis of Witten,
partially proven in \sethi\ .  
The multiplicities and energy spectra of Kaluza-Klein states in a 
frame where their nine dimensional spatial momentum is much smaller than
their mass are exactly the same as the multiplicities and energy 
spectra of massless supergravitons in the IMF.  Thus, the
hypothesis that the Hamiltonian \ham\ has exactly one supermultiplet of
$N$ zero brane threshold bound states for each $N$ guarantees that
the IMF theory has single supergraviton states with the right
multiplicities and spectra.

Multi supergraviton states are discovered by looking at the moduli space
of the quantum mechanics.  In quantum field theory with more than one
space dimension, minima of the bosonic potential correspond to classical
ground states.   There are minima of the Hamiltonian \ham\ corresponding
to \lq\lq spontaneous breakdown \rq\rq of $U(N)$ to any subgroup $U(N_1
) \times \ldots \times U(N_k )$.  These correspond to configurations of
the form
\eqn\vev{X^i = \bigoplus_{s=1}^k r^i I_{N_s \times N_s}.}

 The large number of supersymmetries of the present
system would guarantee that there was an exact quantum ground state of
the theory for each classical expectation value.

In quantum mechanics, the symmetry breaking expectation values $r^i$ are not
frozen variables.  However, if we integrate out all of the other
variables in the system, supersymmetry guarantees that the effective
action for the $\vec{r_s}$ contains no potential terms.  All terms are
at least quadratic in velocities of these coordinates.  We will see that
at large $N$, with all ${N_i \over N}$ finite, and whenever the
separations $| {\bf r_i - r_j}|$ are large, these coordinates are the
slowest variables in our quantum system.  The procedure of integrating
out the rest of the degrees of freedom is thereby justified.
We will continue to use the term moduli space to characterize the space of slow
variables in a Born-Oppenheimer approximation, for these slow variables will always
arise as a consequence of SUSY.  In order to avoid confusion with the moduli space of 
string vacua, we will always use the term {\it background} when referring to the
latter concept. 

We thus seek for solutions of the Schrodinger equation for our $N \times
N$ matrix model in the region of configuration space where all of the
$| {\bf r_i - r_j}|$ are large.  We claim that an approximate solution
is given by a product of SUSY ground state solutions of the $N_i \times
N_i$ matrix problems (the threshold bound state wave functions discussed above), 
multiplied by rapidly falling Gaussian wave
functions of the off diagonal coordinates, and scattering wave functions for
the center of mass coordinates (coefficient of the block unit matrix) of the
individual blocks:
\eqn\scattstatea{\Psi = \psi ({\bf r_1} \ldots
{\bf r_k} ) e^{- \ha | {\bf r_i - r_j}| W_{ij}^{\dagger} W_{ij}}
\bigotimes_{s=1}^k  \psi_B (X_{N_s \times N_s}^i )
}
\eqn\scattstateb{Q_{\alpha}^{N_s \times N_s} \psi_B (X_{N_s \times
N_s}^i ) = 0   }
Here $W_{ij}$ is a generic label for off diagonal matrix elements between
the $N_i$ and $N_j$ blocks.
We claim that the equation for the wave function $\psi ({\bf r_1} \ldots
{\bf r_k} )$ has scattering solutions (Witten's conjecture implies that
there is a single threshold bound state solution as well).  

To justify this form, note that for fixed $| {\bf r_i - r_j}|$, the 
$[X^a , X^b ]^2$ interaction makes the $W_{ij}$ variables into harmonic
oscillators with frequency $| {\bf r_i - r_j}|$.  This is just the
quantum mechanical analog of the Higgs mechanism.  For large
separations, the off diagonal blocks are thus high frequency variables
which should be integrated out by putting them in their (approximately
Gaussian) ground states.  SUSY guarantees that the virtual effects of
these DOF will not induce a Born-Oppenheimer potential for the slow
variables ${\bf r_i}$.  Indeed, with $16$ SUSY generators we have an
even stronger nonrenormalization theorem: the induced effective Lagrangian
begins at quartic order in velocities (or with multifermion terms of the
same \lq\lq supersymmetric dimension \rq\rq).  Dimensional analysis then
shows that the coefficients of the velocity dependent terms fall off as
powers of the separation \dkps\ .  The effective Lagrangian which
governs the behavior of the wave function $\psi$ is thus
\eqn\effham{\sum_{s=1}^k \ha {N_s \over R} {\dot{{\bf r_s}} }^2 + H.O.T.}
where $H.O.T.$ refers to higher powers of velocity and inverse separations.
This clearly describes scattering states of $k$ free particles with a
relativistic eleven dimensional dispersion relation.  The free particle
Hamiltonian is independent of the superpartners of the  ${\bf r_i}$.
This implies a degeneracy of free particle states governed by the
minimal representation of the Clifford algebra
\eqn\cliff{[\Theta_{\alpha} , \Theta_{\beta} ]_+ = 
\delta_{\alpha\beta}}
This has $256$ states.  The $\Theta_{\alpha}$ are in the ${\bf 16}$ of
$SO(9)$, so the states decompose as ${\bf 44 + 84 + 128}$ which is
precisely the spin content of the eleven dimensional supergraviton.

Thus, given the assumption of a threshold bound state in each $N_i$
sector, we can prove the existence, as $\nto$, of the entire Fock space
of SUGRA.  To show that it is indeed a Fock space, we note that the
original $U(N)$ gauge group contains an $S_k$ subgroup which permutes
the $k$ blocks.  This acts like statistics of the multiparticle
states. The connection between spin and statistics follows from the
fact that the fermionic coordinates of the model are spinors of the
rotation group.

It is amusing to imagine an alternative history in which free quantum
field theory was generalized not by adding polynomials in creation and
annihilation operators to the Lagrangian, but by adding new degrees of
freedom to convert the $S_N$ statistics symmetry into a $U(N)$ gauge
theory.  We will see a version of this mechanism working also in the
weakly coupled string limit of the matrix model.  Amusement aside, it is
clear that the whole structure depends sensitively on the existence of
SUSY.  Without SUSY we would have found that the zero point fluctuations
of the high frequency degrees
of freedom induced a linearly rising Born-Oppenheimer potential between
the would be asymptotic particle coordinates.  There would have been no
asymptotic particle states.  In this precise sense, {\it locality and
cluster decomposition are consequences of SUSY in the matrix model}.  It
is important to point out that the crucial requirement is {\it
asymptotic SUSY}.  In order not to disturb cluster decomposition, 
SUSY breaking must be characterized by a finite
energy scale and must not disturb the equality of the term linear in distance
in the frequencies of bosonic and fermionic off diagonal oscillators.
Low energy breaking of SUSY which does not change the coefficients of these
infinite frequencies, is sufficient to guarantee the existence of 
asymptotic states.
The whole discussion is reminiscent of the conditions for absence of a
tachyon in perturbative string theory.   

We end this section by writing a formal expression for the S-matrix of the
finite $N$ system.  It is given\ref\fadeev{L.Fadeev, in {\it Methods in
Field Theory}, Les Houches Lectures 1975, North Holland.} 
by a path integral of the
matrix model action, with asymptotic boundary conditions:
\eqn\asya{X^i (t) \rightarrow \bigoplus_{s=1}^k R\ {(p^s_{\pm} )^i \over N_s}\ t\
I_{N_s \times N_s}}
\eqn\asyb{\Theta_{\alpha} (t)\rightarrow \theta^{\pm}_{\alpha}}
\eqn\asyc{t \rightarrow \pm \infty}
This formula is the analog of the LSZ formula in field theory\foot{
For an alternate approach to the scattering problem, as well as detailed
calculations, see the recent paper \ref\plefka{J. Plefka, A.Waldron, hep-th/9710104.}}.

As a consequence of supersymmetry, 
we know that the system has no stable finite 
energy bound states 
apart from the threshold bound state supergravitons
 we have discussed above.  The boundary conditions
\asya\ - \asyc\ fix the number and quantum numbers of incoming and outgoing 
supergravitons, as long as the threshold bound state wave functions do not vanish at
the origin of the nonmodular coordinates.  The path integral will be equal to the
scattering amplitude multiplied by factors proportional to the bound state wave
function at the origin.  These renormalization factors might diverge or go to zero in
the large $N$ limit, but for finite $N$ the path integral defines a finite unitary
S matrix.  The existence of the large $N$ limit of the S-matrix is 
closely tied up with the nonmanifest Lorentz symmetries.  Indeed, the existence
of individual matrix elements is precisely the statement of longitudinal boost invariance. 
Boosts act to rescale the longitudinal momentum and longitudinal boost invariance means
simply that the matrix element depend only on the ratios ${N_i \over N_k}$ of the
block sizes, in the large $N$ limit.  As a consequence of exact unitarity and energy
momentum conservation, the only disaster which could occur for 
the large N limit of a longitudinally boost invariant system is an infrared catastrophe.
The probability of producing any finite number of of particles from an initial state
with a finite number of particles might go to zero with $N$.  In low energy SUGRA, this does
not happen, essentially because of the constraints of eleven dimensional Lorentz
invariant kinematics.  Thus, it appears plausible that the existence of a finite
nontrivial scattering matrix for finite numbers of particles in the large $N$ limit
is equivalent to Lorentz invariance.  Below we will present evidence that certain
S-matrix elements are indeed finite, and Lorentz invariant.

\subsec{Exhibit M}

The successes of M theory in reproducing and elucidating properties of string vacua
depend in large part on structure which goes beyond that of eleven dimensional SUGRA.
M theory is hypothesized to contain infinite BPS membrane and five brane states.  These 
states have tensions of order the appropriate power of the eleven dimensional Planck scale
and cannot be considered part of low energy SUGRA proper.  However, the behavior of their
low energy excitations and those of their supersymmetrically compactified relatives, is 
largely determined by general properties of quantum mechanics and SUSY.  This information
has led to a large number of highly nontrivial results \mthexamples\ .  The purpose of
the present subsection is to determine whether these states can be
discovered in the matrix model.   

We begin with the membrane, for which the answer to the above question is an unequivocal
and joyous yes.  Indeed, membranes were discovered in matrix models in beautiful work
which predates M theory by almost a decade \ref\membmat{B. de Wit, J.
Hoppe, H. Nicolai, Nucl.Phys. B 305 [FS 23] (1988)
545.}.  Some time before the paper
of \bfss\ Paul Townsend \ref\towns{P.K. Townsend, 
"D-branes {}from M-branes", Phys. Lett. { B373} (1996) 68,
hep-th/9512062 .} 
pointed out the connection between this early work and the
Lagrangian for $D0$ branes written down by Witten.  

This work is well documented in the literature \membmat\ 
, and we will content ourselves with a brief summary and a list of important points.
The key fact is that the algebra of $N\times N$ matrices is generated by a 't Hooft-
Schwinger-Von
Neumann-Weyl pair of conjugate unitary operators $U$ and $V$ satisfying
the relations\foot{The relationship between matrices and membranes was
first explored in this basis by \ref\zachos{D.Fairlie, P.Fletcher,
C.Zachos, Phys. Lett. 218B, (1989), 203, J.Math.Phys. 31, (1990), 1088, 
D.Fairlie, C.Zachos, Phys. Lett. 224B, (1989), 101.}}
\eqn\rela{U^N = V^N = 1}
\eqn\relb{U V = e^{{2\pi i \over N}} VU}
In the limit $\nto$ it is convenient to think of these as $U = e^{iq}$, $V = e^{ip}$ with 
$[q,p] = {2\pi i \over N}$, though of course the operators $q$ and $p$ do not exist for 
finite $N$.   If $A_i = \sum a_i^{mn} U^m V^n$ are large $N$ matrices whose Fourier 
coefficients $a_i^{mn}$ define smooth functions of $p$ and $q$ when the latter are
treated as c numbers, then
\eqn\comm{[A_i , A_j ] \rightarrow {2\pi i \over N} \{ A_i , A_j \}_{PB}}
It is then easy to verify \membmat\ that the matrix model Hamiltonian and SUSY charges 
formally converge to those of the light cone gauge eleven dimensional supermembrane, when
restricted to these configurations.

We will not carry out the full Dirac quantization of the light cone gauge supermembrane 
here, since that is well treated in the early literature.  However, a quick, heuristic
treatment of the bosonic membrane may be useful to those readers who are not familiar 
with the membrane literature, and will help us to establish certain important points.
The equations of motion of the area action for membranes
 may be viewed as the current conservation laws for the spacetime 
momentum densities
\eqn\momdens{P^{\mu}_A = {\partial_A x^{\mu}\over \sqrt{g}} }
where $g_{AB} = (\partial_{A} x^{\mu}) (\partial_{B} x_{\mu})$ is the 
metric induced on the world volume by the background Minkowski space.
In lightcone gauge we choose the world volume time equal to the time in some light cone 
frame
\eqn\time{t = x^+}
We can now make a time dependent reparametrization of the spatial world volume coordinates 
which sets 
\eqn\shift{g_{a0} = {\partial x^- \over \partial\sigma^a} +  {\partial x^i \over \partial 
t}  {\partial x^i \over \partial\sigma^a} = 0.}
  This leaves us only time independent reparametrizations as
a residual gauge freedom.  If $G_{ab}$ is the spatial world volume metric, the 
equation for conservation
of longitudinal momentum current becomes:
\eqn\longcons{\partial_t P^+ = \partial_t \sqrt{G\over g_{00}} = 0,}
where $P^+$ is the longitudinal momentum density.  Since the longitudinal 
momentum density is time independent, we can do a reparametrization at the initial time
which makes it uniform on the world volume, and this will be preserved by the dynamics.
We are left finally with time independent, area preserving diffeomorphisms as gauge 
symmetries.  Note also that, as a consequence of the gauge conditions, $G_{ab}$ depends
only on derivatives of the transverse membrane coordinates $x^i$.

As a consequence of these choices, the equation of motion for the transverse coordinates
reads
\eqn\eom{\partial_t (P^+ {\partial_t x^i}) + \partial_a ({1\over P^+} \epsilon^{ac} 
\epsilon^{bd} \partial_c x^j \partial_d x^j \partial_b x^i) = 0.}
This is the Hamilton equation of the Hamiltonian
\eqn\ham{H = P_- = {1\over P^+} [\ha (P^i )^2 + (\{ x^i , x^j\})^2]}
Here the transverse momentum is $P^i = P^+ \partial_t x^i$ and the Poisson bracket\foot{
This is not the Poisson bracket of the canonical formalism, which is replaced by operator
commutators in the quantum theory.  It is a world volume symplectic structure which is
replaced by matrix commutators for finite $N$.} is
defined by $\{ A, B\} = \epsilon^{ab} \partial_a A \partial_b B$.
The residual area preserving diffeomorphism invariance allows us to
choose $P^+$ to be constant over the membrane at the initial time, and 
the equations of motion guarantee that this is preserved in time.  $P^+$
is then identified with $N/R$, the longitudinal momentum.  For the
details of these constructions, we again refer the reader to the
original paper, \membmat\ .

It is important to realize precisely what is and is not established by this result.
What {\it is} definitely established is the existence in the matrix model spectrum, of 
metastable states which propagate for a time as large semiclassical membranes.  To establish
this, one considers classical initial conditions for the large $N$ matrix model, for which
all phase space variables belong to the class of operators satisfying \comm\ .  One 
further requires that the membrane configurations defined by these initial conditions
are smooth on scales larger than the eleven dimensional Planck length.  It is then easy 
to verify that by making $N$ sufficiently large and the membrane sufficiently smooth, the 
classical matrix solution will track the classical membrane solution for an
arbitrarily long time.  It also appears that in the same limits, the quantum corrections to
the classical motion are under control although this claim definitely needs work.  In
particular, it is clear that the nature of the quantum corrections depends crucially on
SUSY.   The classical motion will exhibit phenomena associated with the flat directions
we have described above in our discussion of the supergraviton Fock space.
In membrane language, the classical potential energy vanishes for membranes of zero
area.  There is thus an instability in which a single large membrane splits into two
large membranes connected by an infinitely thin tube.  Once this happens, the membrane
approximation breaks down and we must deal with the full space of large $N$ matrices.
The persistence of these flat directions in the quantum theory requires SUSY.

Indeed, I believe that quantum membrane excitations of the large $N$ matrix
model will only exist in the SUSY version of the model.  
Membranes are states with classical energies
of order $1/N$.  Standard large $N$ scaling arguments, combined with
dimensional analysis (see the Appendix of \bfss\ )  lead one to the estimate $E \sim N^{1/3}$ for 
typical energy scales in the bosonic matrix quantum mechanics.   The quantum corrections to
the classical membrane excitation of the large $N$ bosonic matrix model completely
dominate its energetics and probably qualitatively change the nature of the state.

One thing that is clear about the quantum corrections is that they have nothing to
do with the quantum correction in the nonrenormalizable field theory defined by
the membrane action.   We can restrict our classical initial matrix data to resemble 
membranes and with appropriate smoothness conditions the configuration will propagate as
a membrane for a long time.  However, the quantum corrections involve a path integral
over all configurations of the matrices, including those which do not satisfy \comm\ .
The quantum large $N$ Matrix Theory is not just a regulator of the membrane action with 
a cutoff going to infinity with $N$.  It has other degrees of freedom which cannot be 
described as membranes even at large $N$.
In particular (though this by no means exhausts the non-membrany configurations of the 
matrix model), the matrix model clearly contains configurations
containing {\it an arbitary number
of membranes}.  These are block diagonal matrices with each block containing a finite 
fraction of the total $N$, and satisfying \comm\ .   The existence of the continuous
spectrum implied by these block diagonal configurations was first pointed out in
\ref\dln{B. de Wit, M.
Luscher and H. Nicolai, Nucl. Phys. B 320 (1989) 135.}.

The approach to membranes described here emphasizes the connection to toroidal membranes.
The basis for large $N$ matrices which we have chosen, is in one to one correspondence
with the Fourier modes on a torus.  The finite $N$ system has been described by
 mathematicians as the noncommutative or fuzzy torus.  
 In fact, one can find bases corresponding to a complete set of functions on any Riemann
surface\ref\dhnbars{de Wit, Hoppe, Nicolai, {\it op. cit.}, I.Bars, 
hep-th/9706177, and references therein.} .  
The general idea is to solve the quantum mechanics problem of
a charged particle on a Riemann surface pierced by a constant magnetic field.  This system
has a finite number of quantum states, which can be parametrized by the guiding center
coordinates of Larmor orbits.  In quantum mechanics, these coordinates take on only
a finite number of values.  As the magnetic field is taken to infinity, the system becomes
classical and the guiding center coordinates become coordinates on the classical Riemann
surface.   
What is most remarkable about this is that for finite $N$ we can choose any basis we wish 
in the space of matrices.  They are all equivalent.  Thus, the notion of membrane topology
only appears as an artifact of the large $N$ limit.

\subsec{Scattering}

We have described above a general recipe for the scattering matrix in Matrix Theory.
In this section we will describe some calculations of scattering amplitudes in a dual
expansion in powers of energy and inverse transverse separation.   The basic idea 
is to exploit the Born-Oppenheimer separation of energy scales which occurs when
transverse separations are large.  Off diagonal degrees of freedom between blocks
acquire infinite frequencies when the separations become large.  The coefficient of
the unit matrix in each block, the center of mass of the block,
 interacts with the other degrees of freedom in
the block only via the mediation of these off diagonal \lq\lq W bosons \rq\rq .
Finally, the internal block degrees of freedom are supposed to be put into the
wave function of some composite excitation (graviton or brane).  We will present evidence
below that the internal excitation energies in these composite wave functions are
, even at large N, parametrically larger than the energies associated with motion of
the centers of mass of blocks of size $N$ with finite transverse momentum.
Thus the center of mass coordinates are the slowest variables in the system and
we can imagine computing scattering amplitudes from an effective Lagrangian which
includes only these variables.

To date, all calculations have relied on terms in the effective action which come
from integrating out W bosons at one or two loops.  It is important to understand that
the applicability of perturbation theory to these calculations is a consequence of the
large W boson frequencies.  The coupling in the quantum mechanics is {\it relevant} 
so high frequency loops can be calculated perturbatively.  The perturbation parameter
is $({\lp \over r})^3$ where $r$ is a transverse separation.  In most processes which
have been studied to date, effects due to the internal block wave functions, are higher
order corrections.  The exception is the calculation of
\ref\ganram{O.J. Ganor, R.Gopakumar, S.Ramgoolam, hep-th/9705188. }, which fortuitously
depended only on the matrix element of the canonical commutation relations in the bound
state wave function.  It would be extremely interesting to develop a systematic formalism
for computing wave function corrections to scattering amplitudes.  Since the center of
mass coordinates interact with the internal variables only via mediation of the
heavy W bosons, it should be possible to use Operator Product Expansions in the
quantum mechanics to express amplitudes up to a given order in energy and transverse
distance in terms of the matrix elements of a finite set of operators.

Almost all of the calculations which have been done involve zero longitudinal
momentum transfer.  The reason for this should be obvious.  A process involving
nonzero momentum transfer requires a different block decomposition of the matrices in the
initial and final states.  It is not obvious how to formulate this process in a
manner which is approximately independent of the structure of the wave function.
In a beautiful paper, Polchinski and Pouliot \ref\pp{J.Polchinski,
P.Pouliot, hep-th/9704029.} were able to do a computation with
nonzero longitudinal momentum transfer between membranes.  The membrane is a semiclassical
excitation of the matrix model, and thus its wave function, unlike that of the
graviton is essentially known.  We will describe only the original \bfss\ calculation of supergraviton
scattering.  Other calculations, which provide extensive evidence for Matrix Theory,
will have to be omitted for lack of space.  We refer the reader to the literature
\ref\scattering{S.D.Mathur, G.Lifschytz, hep-th/9612087; G.Lifschytz,
hep-th/9612223, O.Aharony, M.Berkooz, Nucl.Phys. B491 (1997) 184,
hep-th/9611215; K.Becker, M.Becker, J.Polch\break inski, A.Tseytlin,
Phys.Rev. D56 (1997) 3174, hep-th/9706072; K.Becker, M.Becker,
hep-th/9705091; J.Harvey, hep-th/9706039; P.Berglund, D.Minic,
hep-th/9708063; R.Gopa-kumar, S.Ramgoolam, hep-th/9708022;
J.M.Pierre, hep-th/9705110; I.Chepel-ev, A.A.Tseytlin, Phys.Rev. D56
(1997) 3672, hep-th/9704127;  }.

The amplitude for supergraviton scattering can be calculated by a simple extension
of the zero brane scattering calculation performed by
\ref\bachas{C.Bachas, Phys.Lett. B374 (1996) 37, hep-th/9511043.} and \dkps\ .
By the power counting argument described above, the leading order contribution at large
transverse distance to the term in the effective action with a fixed power of the relative
velocity is given by a one loop diagram.  For supergravitons of $N_1$ and
$N_2$ units of longitudinal momentum, the two boundary loops in the diagram give a factor
of $N_1 N_2$ relative to the zerobrane calculation.  We also recall that the amplitude
for the particular initial and final spin states defined by the boundary state
of \bachas\ depends only on the relative velocity ${\bf v_1 - v_2 } \equiv {\bf v}$ of the
gravitons.  As a consequence of nonrenormalization theorems the interaction correction 
to the effective Lagrangian
begins at order $({\bf v}^2)^2 \equiv v^4$.  

Apart from the factor of $N_1 N_2$ explained above, 
the calculation of the effective Lagrangian was performed in \dkps\ .
It gives
\eqn\leffn{L = {N_1{\dot r(1)}^2 \over 2R}+{N_2{\dot r(2)}^2 \over 2R}
- A N_1N_2{[\dot r(1)-\dot r(2)]^4\over {R^3 {(r(1) - r(2))}^7}}}
The coefficient $A$ was calculated in \dkps\ .  For our purposes it
is sufficient to know that this Lagrangian exactly reproduces the
effect of single graviton exchange between D0 branes in ten dimensions. 
This tells us that the amplitude described below is in fact the correctly
normalized {\it eleven} dimensional amplitude for zero longitudinal
momentum exhange in tree level SUGRA.

Assuming the distances are large and the velocity small, the effective
Hamiltonian is 
\eqn\hef{
H_{eff}= { p_{\perp}(1)^2\over 2\p(1)}+{ p_{\perp}(2)^2\over 2\p(2)}
         +A\left[{p_{\perp}(1) \over \p(1)}-{p_{\perp}(2) \over \p(2)}\right]^4
{\p(1)\p(2) \over r^7 R}}
where $r$ is now used to denote the transverse separation.  Treating the
perturbation in Born approximation we can compute the leading order
scattering amplitude at large impact parameter and zero longitudinal
momentum.   It corresponds precisely to the amplitude calculated in
eleven dimensional SUGRA.

We can also use the effective Hamiltonian to derive various interesting facts about the
bound state wave function of a supergraviton.  Let us examine the wave function 
of a graviton of momentum $N$ 
along a flat direction in configuration space corresponding to a pair of clusters
of momenta $N_1$ and $N_2$ separated by a large distance $r$.  The effective Hamiltonian
for the relative coordinate is
\eqn\clusterham{{p^2 \over \mu} + ({N_1 N_2 \over \mu^4}) {p^4 \over r^7}}
where $\mu$ is the reduced mass, ${N_1 N_2 \over {N_1 + N_2}}$.  Scaling this 
Hamiltonian, we find that the typical distance scale in this portion of configuration
space is $r_m \sim ({({N_1 + N_2})^3 \over N_1^2 N_2^2})^{1/9}$, while the typical
velocity is $v_m \sim {1 \over \mu r_m} \sim (N_1 + N_2)^{2/3} ({N_1 N_2})^{-7/9}$.
The typical energy scale for internal motions is $\mu v_m^2$. 
As $N$ gets large the system thus has a continuous range of internal scales.  As in
perturbative string theory, the longest distance scale $\sim N^{1/9}$ is associated
with single parton excitations, with typical energy scale $\sim N^{-2/9}$.  Notice that 
all of these internal velocities get small, thus justifying various approximations
we have made above.  However, even the smallest internal velocity, $\sim N^{ - 8/9}$
characteristic of two clusters with finite fractions of the longitudinal
momentum, is larger than the scale of motions of free particles , $\sim 1/N$.  
This is the justification for treating the coordinates of the centers of mass as
the slow variables in the Born-Oppenheimer approximation.

These estimates also prove that the Bekenstein bound is satisfied 
in our system.  For suppose that the size of the system grew more slowly with
$N$ than $N^{1/9}$.  Then our analysis of a single parton separated from the rest of
the system would show that there is a piece of the wave function with scale $N^{1/9}$
contradicting the assumption.  The analysis suggests that in fact the Bekenstein bound
is saturated but a more sophisticated calculation is necessary to prove this.

We are again faced with the paradox of the introduction: How can systems whose
size grows with $N$ in this fashion have $N$ independent scattering amplitudes
as required by longitudinal boost invariance?  Our results to date only supply clues
to the answer.  We have seen that to leading order in the long distance expansion, the
zero longitudinal momentum transfer scattering amplitudes are in fact Lorentz invariant.
This depended crucially on SUSY.  The large parton clouds are slowly moving BPS
particles, and do not interact with each other significantly.  In addition, we have
seen that the internal structure of the bound state is characterized by a multitude of
length and energy scales which scale as different powers of $N$.  Perhaps this is
a clue to the way in which the bound state structure becomes oblivious to rescaling
of $N$ in the large $N$ limit.   Further evidence of Lorentz invariance of the theory
comes from the numerous brane scattering calculations described in \scattering\ (and
in the derivation of string theory which we will provide in the next section).
Perhaps the most striking of these is the calculation of \pp\ which includes
longitudinal momentum transfer.

\newsec{\bf Compactification}

We now turn to the problem of compactifying the matrix model and begin
to deal with the apparent necessity of introducing new degrees of
freedom to describe the compact theory.  One of the basic ideas which
leads to a successful description of compactification on $\td$
is to look for representations of the configuration space variables
satisfying
\eqn\torus{X^a + 2\pi R^a_i = U_i^{\dagger} X^a U_i \qquad a = 1\ldots d}
This equation says that shifting the dynamical variables $X^a$ by the
lattice which defines $\td$ is equivalent to a unitary transformation.

A very general representation of this requirement is achieved by
choosing the $X^a$ to be covariant derivatives in a $U(M)$ gauge bundle on a
dual torus $\tdd$ defined by the shifts
\eqn\dualtorus{\sigma^a \rightarrow \sigma^a + E^a_i \qquad 2\pi E^a_i
R^a_j = \delta_{ij}}
\eqn\dualtorusb{X^a = {1\over i}{\partial\over \partial\sigma^a}
I_{M\times M} - A_a (\sigma )}

If this expression is inserted into the matrix model Hamiltonian
we obtain the Hamiltonian for maximally supersymmetric Super
Yang Mills Theory compactified on $\tdd$ .  The coordinates in the
compact directions are effectively replaced by gauge potentials, while
the noncompact coordinates are Higgs fields.  It is clear that in the
limit in which all of the radii of $\td$ become large, the dual radii
become small. We can do a Kaluza-Klein reduction of the degrees of
freedom and obtain the original eleven dimensional matrix model.  It is
then clear that we must take the $M\rightarrow\infty$ limit.

The value of the $\symd$ coupling is best determined by computing the
energy of a BPS state and comparing it to known results from string
theory.  The virtue of this determination is that it does not require us
to solve $\symd$ nor to believe that it is the complete theory in all
cases.  It is sufficient that the correct theory reduces to
semiclassical $\symd$ in
some limit.  In this case we can calculate the BPS energy exactly from
classical $\symd$ dynamics.    We will perform such calculations below.
For now it will suffice to know that $g_{SYM}^2 \sim \prod {1\over R_a}$
(here and henceforth we restrict attention to rectilinear tori with
radii $R_a$).  To see this note that the longitudinal momentum is given
by the trace of the identity operator, which involves an integral over
the dual torus.  This parameter should be independent of the background,
which means that the trace should be normalized by dividing by the
volume of the dual torus.  This normalization factor then appears in the
conventionally defined SYM coupling.

In \bfss\ and \ref\taylor{W.Taylor IV, Phys. Lett. B394 (1997) 283,
hep-th/9611042. }\ another derivation of the SYM prescription for
compactification was given.  The idea was to study zero branes in 
weakly coupled IIA string theory compactified on a torus\foot{This idea
was mentioned to various authors of \bfss\ by N. Seiberg, and
independently by E. and H. Verlinde at the Santa Barbara Strings 96
meeting and at the Aspen Workshop on Duality.  The present author did
not understand at the time that this gave a prescription identical to the more
abstract proposal of the previous paragraph.  As usual, progress could
have been made more easily of we had listened more closely to our
colleagues.} .   The nonzero momentum modes of $\symd$ arise in this
context as the winding modes of open IIA strings ending on the zero
branes.  T duality tells us that there is a more transparent
presentation of the dynamics of this system in which the zero branes are
replaced by $d$-branes and the winding modes become momentum modes.  
In this way, the derivation of the compactified theory follows precisely
the prescription of the infinite volume derivation.  This approach also
makes it obvious that new degrees of freedom are being added in the
compactified theory.   

Before going on to applications of this prescription for
compactification, and the ultimate necessity of replacing it by
something more general, I would like to present a suggestion that in
fact the full set of degrees of freedom of the system are indeed present
in the original matrix model, or some simple generalization of it.
This contradicts the philosophy guiding the bulk of this review, but the
theory is poorly understood at the moment, so alternative lines of
thought should not be buried under the rug.  The point is, that 
the expressions \dualtorusb\ for the coordinates in the compactified
theory, are operators in a Hilbert space, and can therefore be
approximated by finite matrices.  Thus one might conjecture that in the
large $N$ limit, the configuration space of the finite $N$ matrix model
breaks up into sectors which do not interact with each other (like
superselection sectors in infinite volume field theory) and that
\dualtorusb\ represents one of those sectors.  The failure of the SYM
prescription above $d=4$ might be viewed simply as the failure of
\dualtorusb\ to include all degrees of freedom in the appropriate
sector.  Certainly, up to $d=3$ we can interpret the SYM prescription 
as a restriction of the full matrix model to a subset of its degrees of
freedom.   We simply approximate the derivative operators by ({\it
e.g.}) $(2P + 1)$ dimensional diagonal 
matrices with integer eigenvalues and the
functions of $\sigma$ by functions of the unitary shift operators which
cyclically permute the eigenvalues.  Choosing $N = (2P + 1)M$ we can
embed the truncated SYM theory into the $U(N)$ matrix model.  Readers
familiar with the Eguchi-Kawai reduction of large $N$ gauge theories
will find this sort of procedure natural\ref\ek{Phys.Rev.Lett. 48 (1982)
1063.}.  

What has not been shown is that the restriction to a particular sector
occurs dynamically in the matrix model.  Advocates of this point of view
would optimistically propose that the dynamics not only segregates the
SYM theory for $d \leq 3$ but also chooses the correct set of degrees of
freedom for more complex compactifications.  The present author is
agnostic about the correctness of this line of thought.  Demonstration
of its validity certainly seems more difficult than other approaches to
the subject of compactification, which we will follow for the rest of
this review.   

In the next section we will show that the SYM
prescription reproduces toroidally compactified Type IIA string theory
for general $d$.  This implies that the eventual replacement of the SYM
theory for $d > 3$ must at least have a limit which corresponds to the
dimensional reduction of $SYM_{d+1}$ to $1+1$ dimensions.   In the next section
 we demonstrate that the SYM prescription for compactification on
$\ttwo$ reproduces the expected duality symmetries of M theory.  In
particular, we identify the Aspinwall-Schwarz limit of vanishing
toroidal area, in which the theory reduces to Type IIB string theory.
Our dynamical approach to the problem enables us to verify the $SO(8)$
rotation invariance between the seven noncompact momenta and the one
which arises from the winding number of membranes.  This invariance was
completely mysterious in previous discussions of this limit.  We are
also able to explicitly exhibit D string configurations of the model and
to make some general remarks about scattering amplitudes.
We then discuss compactification of three dimensions and
exhibit the expected duality group.  Moving on to four dimensions
we 
show that new degrees of freedom, corresponding to five branes wrapped around 
the longitudinal and torus directions, must be added to the theory.  The result
is a previously discovered $5 + 1$ dimensional superconformal field theory.
Compactification on a five torus seems to lead to a new theory which
cannot be described as a quantum field theory, while the six torus is
still something of a mystery.

\newsec{\bf IIA Strings from Matrices}
\subsec{Normalizations}

Before beginning the main work of this section we fix the parameters in
our SYM theory.  We do this by computing the energies of BPS states.
Such computations should be valid even if SYM is only an approximate
description of the theory in some range of parameters.

The first BPS charge which we investigate is the Kaluza-Klein momentum.
We consider a state with one unit of Kaluza-Klein momentum and one unit
of longitudinal momentum.
The lowest state with these quantum numbers has IMF energy ${R\over 2 R_i^2} $
where $R_i$
is the radius of the $ith$ cycle of the torus in M theory. We define
$L_i$ to be the circumference of this cycle.
 The SYM action is
\eqn\symlag{ {1\over g_{SYM}^2} \int d^d \sigma\ \tr [ {\dot{A^a}}^2 - 
{1\over 4} F_{ij}^2 + \ha \dot{X^i}^2 - ({\bf D_{\sigma}} X^i )^2 +
{1\over 4} {[X^i , X^j ]}^2 ] + fermions}

Compactified momentum around a given circle is identified with the
electric flux around that circle. For a state with a single unit of
longitudinal momentum this is the $U(1)$ flux
associated with a $1 \times 1$ block in the $U(N)$ SYM theory. 
This quantity appears as a BPS
central charge of the SUSY algebra of SYM theory.  The associated energy
for a unit $U(1)$ flux is ${g_{SYM}^2  \Sigma_i^2 \over V_{SYM}} $, 
where $\Sigma_i$ is
the circumference of the SYM cycle, and $V_{SYM}$ the volume of the SYM
torus.   Thus we conclude
that ${g_{SYM}^2 \Sigma_i^2 \over V_{SYM}} = {(2\pi )^2 R \over L_i^2}$.

We can also study membranes wrapped around the longitudinal direction
and one of the transverse directions.  The corresponding quantum number
is the momentum on the Yang-Mills torus.  This is analogous to string
theory, where $L_0 - \bar{L_0}$, the world sheet momentum in light cone
gauge, is set equal to the winding number of longitudinal strings by the
Virasoro condition.  Indeed, in the matrix model, the Yang Mills
momentum should be considered a gauge generator, for it generates
unitary transformations on the configuration space of the model.
The fields in the (classical) SYM theory should be thought of as
operators on a Hilbert space of $M$ vector valued functions.  They are
infinite dimensional matrices.   Translations on the SYM torus (not
spacetime translations) are unitary transformations on this Hilbert
space, which preserve the trace operation, $\int d^d \sigma Tr_M $.  
They are analogs of the $U(N)$ gauge transformations of the finite $N$
matrix model.   

The energy of the lowest lying state carrying momentum in the $ith$
SYM direction is
precisely ${2\pi \over \Sigma_i}$.  From the M theory point of view
these states are longitudinally wrapped membranes with energy
${RL \over 2\pi\lp^3}$.   Combining the D0 brane and longitudinally
wrapped membrane formulae, we obtain the relation between the SYM
coupling and radii, and the parameters of M theory:
\eqn\symMrela{\Sigma_i = {4\pi^2 \lp^3 \over L_i R}}
\eqn\symMrelb{g_{SYM}^2 = {R^3 V_{SYM} \over 4\pi^2\lp^6 }}
Note that the dimensionless ratio $g_{SYM}^2 V_{SYM}^{- {d-3 \over d}}$,
which for tori with all radii similar measures the effective coupling at
the size of the SYM torus, is independent of $R$.

With these definitions we can go on to study other BPS states of M
theory in the SYM language, 
\ref\grt{O.J.Ganor, S.Ramgoolam,W.Taylor IV, Nucl. Phys. B492 (1997)
191, hep-th/9611202.}\ ,
\ref\bss{T.Banks, N.Seiberg, S.Shenker,Nucl.Phys. B490 (1997) 91,
hep-th/9612157.}\ ,  
\ref\shrink{F.Halyo, W.Fischler, A.Rajaraman, L.Susskind, 
 hep-th/9703102.}.
Transversely wrapped membranes are associated with magnetic fluxes.
On a  torus with four or more dimensions, we associate
instanton number with the charge of fivebranes wrapped around the
longitudinal direction and a transverse 4 cycle.   The
energy formulae agree with M theory expectations, including the correct
value of the five brane tension.
 The existence of these extra finite
longitudinal charges will turn out to be crucial below.  Finally, we
note that the SYM prescription gives no apparent candidate for the
wrapped transverse fivebrane.  This is one of the clues which suggests
that the SYM prescription is missing something important.

\subsec{How M Theory Copes With the Unbearable Lightness of String}

According to the folklore M theory becomes IIA string
theory when it is compactified on a circle of radius $R_1$ much smaller than
$\lp$ .  Membranes wrapped around the small circle become strings with
tension of order $R_1 \lp^{-3}$.  These lightest objects in the theory
are supposed to become weakly coupled.  The folklore gives no hint as to
how the weak coupling arises.  Our task in this section is to show that
this scenario is realized dynamically in the SYM prescription for Matrix
Theory compactification.

The first step is to note that in the IIA string limit, all other
compactified dimensions are supposed to be of order the string length,
which is much bigger than $\lp$.  This means that the SYM torus has one
large radius, of order $R_1^{-1}$ with all other radii of order
$\sqrt{R_1}$ (everything in $\lp$ units).  Thus, we can perform a
Kaluza-Klein reduction of the SYM theory, turning it into a $1+1$
dimensional field theory.  The degrees of freedom which are being
integrated out in this procedure have energies of order $R_1^{-3/2}$ in
string units, when the radii of the other compactified dimensions are of
order one in string units.  It is also important to note that our
analysis will be valid for any high energy theory which reduces to SYM
theory in the stringy regime.  So far this appears to be the case for
all proposals which replace the nonrenormalizable SYM theory in $d > 3$
with a well defined Hamiltonian.  The new degrees of freedom which
appear in these models can be identified with branes wrapped around
large cycles of $\td$ and their energy is very large in the stringy
limit. 

{\it It is extremely important at this point that the maximally
supersymmetric SYM theory is uniquely defined by its symmetries and that
there is a nonrenormalization theorem for the SYM coupling}.  As a
consequence, we know that the low energy effective Lagrangian is just
the $1+1$ dimensional SYM theory obtained by classical dimensional
reduction.   This theory lives on a circle with radius $\sim R_1^{-1}$.  It
thus contains very low energy states in the IIA limit of $R_1
\rightarrow 0$.  To isolate the physics of these states we rescale the
coordinate to run from zero to $2\pi$, and simultaneously rescale the
time (and Hamiltonian)
 and the eight transverse coordinates so that the quadratic terms in
their Lagrangian have coefficients of order one.  This corresponds to a
passage from Planck units to string units.  We will exhibit only the
bosonic part of the rescaled Lagrangian since the fermionic terms follow
by supersymmetry.  

Before we do so, we make some remarks about the gauge fields.  $d-1$ of
the eight transverse coordinates arise as Wilson lines of the $d+1$
dimensional gauge theory around the large (in spacetime) 
 compact dimensions.  They have
periodicities $2\pi R_i$, $2\leq i \leq d$.  In the $R_1 \rightarrow 0$
limit, this is the only remnant of the compactness of the large
dimensions.  The gauge potential in the $1$ direction is another beast
entirely. When we make the Kaluza-Klein reduction the formulae relating
the gauge coupling and the volume are such that the conventional $1+1$
dimensional coupling is simply $R_1^{-1}$ with no dependence on the
other radii.  As a consequence, in the IIA limit the gauge theory
becomes very strongly coupled.  There is something to be learned here
about the folkloric picture of the IIA string as a wrapped membrane.
When the $R_1$ circle is large, the variable $A_1$ is semiclassical and
plays the role of a coordinate in the eleventh dimension.  In the IIA
limit however this coordinate is a rapidly fluctuating quantum variable
(indeed its canonical conjugate is approximately diagonal in the ground
state), and simple geometrical pictures involving the eleventh dimension
are completely false.   
The success of the folkloric predictions is a consequence of their
strict adherence to the rule of calculating only BPS quantities.  These
can be understood in a limit of the parameter space in which
semiclassical reasoning is applicable, and the resulting formulae are
valid outside the semiclassical regime.  It is wise however to refrain
from attributing too much reality to the semiclassical picture outside
its range of validity.

The rescaled bosonic Hamiltonian is
\eqn\resclag{H = R \int d\sigma \bigl[ R_1^{-3} E^2 + (P^i )^2 +
(D_{\sigma} X^i)^2 - R_1^{-3} [X^i , X^j ]^2 \bigr]}
The index $i$ runs over the $8$ remaining transverse dimensions, some of
which are compactified.  The coordinates in the compactified dimensions
are really SYM vector potentials, but in the Kaluza-Klein limit which we
are taking the only remnant of the SYM structure is that the
corresponding $X^i$ variables are compact.  They represent Wilson lines
around the compactified SYM directions.
We see that as $R_1 \rightarrow 0$, we are forced onto the moduli space of 
the SYM theory.  This is the space of commuting matrices. Equivalently
it can be described as the space of diagonal matrices modded out by the
Weyl group of the gauge group.  The Weyl group is the semidirect product
of $S_N$ with $T_{(d-1)N}$, the group of integer shifts of a $(d-1)N$
dimensional Euclidean space.
We will refer to the field theory on such a target space as a symmetric product
orbifold theory.  The second factor , $T_{(d-1)dN}$, in the orbifold group, arises because the
$d-1$ compactified coordinates are Wilson lines of the gauge group.  Thus these coordinates 
lie in the Cartan torus, that is, ${\bf R}^{(d-1)N}$  modded out by the group of shifts $T_{(d-1)N}$, 
rather than ${\bf R}^{(d-1)N}$ itself.   

The nonrenormalization theorem for theories with $16$ supercharges tells us that the 
free lagrangian on this orbifold target space is the unique dimension two operator with
the symmetries of the underlying SYM theory.  We will see in a moment that the 
leading irrelevant operator has dimension $3$.  The fact that the effective theory is
free in the $R_1 \rightarrow 0$ limit, is a derivation of one of the central tenets of
string duality ({\it viz.} the existence of an eleven dimensional quantum theory whose
compactification on a zero radius circle gives free string theory) from Matrix Theory.
To complete the derivation, we must show that the spectrum of the symmetric product
orbifold theory is equivalent to that of string field theory.  The central physics issue
here was first pointed out by Motl\ref\motlI{L.Motl, hep-th/9701025.} , although the mathematical framework
had appeared previously in black hole physics and other places
\ref\msetc{J.Maldacena, L.Susskind, Nucl.Phys. B475 (1996) 679, hep-th/9604042; S.Das,
S.D.Ma\break thur, Nucl.Phys. B478 (1996) 561, hep-th/9606185 ; R.Dijkgraaf,
E.Verlinde, H.Verl- inde, Nucl.Phys. B486 (1997) 89, hep-th/9604055 ; R.Dijkgraaf, G.Moore, E.Verlinde, H.Verlinde,
Comm. Math. Phys. 185 (1997) 197, hep-th/9608096; C.Vafa, E.Witten,
Nucl. Phys. B431 (1994) 3, hep-th/9408074 ; }.  
It was rephrased in the language of gauge theory moduli spaces by
\ref\bs{T.Banks, N.Seiberg, Nucl.Phys. B497 (1997) 41, hep-th/9702187.}.  This
was done independently by \ref\dvv{R.Dijkgraaf, E.Verlinde, H.Verlinde, hep-th/9703030.} , who pointed out the origin of the Virasoro
conditions and showed that the leading irrelevant operator reproduced the light cone string
vertex.

The central point is that, as a consequence of the $S_N$ orbifold, the individual
diagonal matrix elements do not have to be periodic with period $2\pi$.  Rather, we can
have twisted sectors of the orbifold QFT.  These sectors correspond to the conjugacy
classes of the orbifold group.  A general permutation is conjugate to a product of
commuting cycles.  As a consequence of the semi-direct product structure of the group, a 
general permutation times a general shift is conjugate to the same permutation times
a shift which is proportional to the unit matrix in each block corresponding to a single
cycle of the permutation.   Let $(k_1 \ldots k_n ), \sum k_i = N$, be the cycle lengths
of the permutation in a particular twisted sector and let $2\pi R_a (m_1^a \ldots m_n^a)$
be the shifts in this sector for the $a$th compactified coordinate.
Then the twisted boundary condition is solved by a diagonal matrix which breaks up into
$n$ blocks.  In the $i$th block, the diagonal matrix is $diag [x_i^a (\sigma ), x_i^a (\sigma + 2\pi ), \ldots x_i^a (\sigma + 2 \pi (k_i -1) )] $.  The variable $\sigma$ runs from $0$ to
$2\pi$.  $x_i^a$ satisfies $x_i^a (s + 2\pi k_i) = x_i^a (s) + 2 \pi R_a m_i$, and the
winding numbers $m_i^a$ vanish in the noncompact directions.   There are similar formulae
for the fermionic variables.  The correspondence  with the Fock space of Type IIA string
field theory is immediately obvious.  The Lagrangian reduces on these configurations
to $n$ copies of the IIA Green-Schwarz Lagrangian, with the longitudinal momentum of
the $i$th string equal to $k_i / R$.  If $k_i$ is proportional to $N$ in the large $N$ 
limit, then these states have energies of order $1/N$.
Note that the prescription of one sector for each conjugacy class of the orbifold group
automatically gives us one winding number for each individual string (and each compact
direction).  Naively, one might have imagined that one had a winding number for each
individual {\it eigenvalue} but these states are just gauge copies of the ones we have
exhibited.

The Virasoro condition is derived, as shown by \dvv , by imposing the $\prod Z_{k_i}$
gauge conditions on the states.   This actually imposes the more general condition
$L_0 - \bar{L_0} = W$, where $W$ is the winding number around the
compact longitudinal direction.  Recall that in conventional light cone string theory,
the Virasoro condition is obtained by integrating the derivative of the longitudinal
coordinate. It is worth noting for later use that the momentum in
the underlying SYM theory is also interpretable as a longitudinal winding number.
It appears in the SUSY algebra \bss\ in the place appropriate for the wrapping number
of a membrane around the torus formed by the longitudinal circle and the small
circle which defines the string coupling.   

From the point of view of SYM theory, the free string limit is the limit of strong coupling
and it is difficult to make explicit calculations of the effective Lagrangian on the
moduli space.  To derive the existence of the free string limit we have used the
method of effective field theory - symmetries completely determine the lowest dimension
effective Lagrangian.  \dvv went further, and showed that the leading correction to
free string theory was also determined essentially by symmetries.  In order to correspond
to a string interaction, and not simply a modification of free string propagation, the
required operator must permute the eigenvalues.  It is thus a twist field of the orbifold.
The lowest dimension twist operator exchanges a single pair of eigenvalues.  However, in
order to be invariant under $SO(8)$ and under SUSY, it must exchange the eigenvalues of
all of the fermionic and bosonic matrices in a single $2 \times 2$ block.
If we define the sum and difference operators $Z_{\pm} = Z_1 \pm Z_2$ for both bosonic and
fermionic coordinates, then we are discussing an operator contructed only out of $Z_-$.
In terms of these variables, the eigenvalue exchage is simply a $Z_2$ reflection, so we can
use our knowledge of the conformal field theory of the simplest of all orbifolds.

Twist operators are defined by the OPE
\eqn\bostwist{\partial x^i_- (z) \tau (0 ) \sim z^{-\ha} \tau^i (0)}
for the bosonic fields, and
\eqn\fermtwist{\theta^{\alpha}_- (z) \Sigma^i (0) \sim z^{-\ha}
\gamma^i_{\alpha\dot{\alpha}} \Sigma^{\dot{\alpha}} (0)}
\eqn\fermtwistb{\theta^{\alpha}_- (z) \Sigma^{\dot{\alpha}} (0) \sim z^{-\ha}
\gamma^i_{\alpha\dot{\alpha}} \Sigma^i (0)}

The operator $\tau$ is the product of the twist operators for the $8$ individual bosons.
It has dimension $\ha$.
The operator $\tau^i$ has dimension 1, transforms as an $SO(8)$ vector, and has
a square root branch point OPE with all the left moving bosonic currents.
The operator $\Sigma^i$ is the product of spin operators for the $8$ left moving 
Green-Schwarz fermions.  It is the light cone Ramond Neveu Schwarz fermion field, and
has dimension $\ha$.   The operator $\tau^i \Sigma^i$ is thus $SO(8)$ invariant.
It is in fact invariant under the left moving SUSY's.  To demonstrate this we need an
identity proven in \dvv\ 
\eqn\dvvident{[G_{-\ha}^{\dot{\alpha}} , \tau \Sigma^{\dot{\beta}} ] + 
[G_{-\ha}^{\dot{\beta}} , \tau \Sigma^{\dot{\alpha}}]  =
\delta^{\dot{\alpha} \dot{\beta}} \tau^i \Sigma^i .}

In principle there could have been more complicated $SO(8)$
representations on the right hand side of this equation, but there is a
null vector in the representation of the SUSY algebra provided by the
vertex operators.
It follows that
\eqn\susy{[G_{-\ha}^{\dot{\alpha}},  \tau^i \Sigma^i] = \partial_z(\tau \Sigma^{\dot{\alpha}}) }
 Thus, the operator $\int \tau^i \Sigma^i  \bar{\tau}^i \bar{\Sigma}^i$ is 
a supersymmetric, $SO(8)$ invariant interaction of dimension $3$.
Here, the barred quantities are right moving fields constructed in a manner precisely 
analogous to their left moving unbarred counterparts.  Apart from the free Lagrangian, 
this is in fact the lowest dimension operator which is invariant under the full SUSY and
$SO(8)$ symmetry group.  It is in fact the operator constructed long ago by 
Mandelstam\ref\mandel{S.Mandelstam, Nucl.Phys. B83 (1974) 413,
Prog. Theor.Phys.Suppl. 86 (1986) 163.} , to describe the three string interaction in light cone gauge.

Using the methods of effective field theory, we cannot of course calculate the precise
coefficient in front of this operator.  This would require us to perform a microscopic
calculation in the strongly coupled SYM theory.  Dimensional analysis tells us that the
coefficient in front of this operator is proportional to $1/M$ the mass scale of the
heavy fields which are integrated out.  Referring back to \resclag\ we see that the heavy
off diagonal fields have masses of order $R_1^{-3/2}$.  We conclude that the string
coupling $g_S \sim R_1^{3/2}$, precisely the scaling predicted by Witten \witten\ .

This is a spectacular success for the matrix model, but we should probe carefully to
see how much of the underlying structure it tests.   In particular, the effective field
theory argument appears to depend only weakly on the fact that the underlying theory  
was SYM.  One could imagine additional high energy degrees of freedom that would lead
to the same leading order operator.  The strongest arguments against the existence of
such degrees of freedom are similar to those given above for the matrix quantum mechanics.
The symmetries of the light cone gauge theory are so constraining that it is unlikely
that we will find another set of canonical degrees of freedom and/or another Lagrangian
(remember that a Lagrangian for a complete set of degrees of freedom is local in time
and can always be brought to a form which involves only first time derivatives)
which obeys them.  The derivation of the correct scaling law for the string coupling
reinforces this conclusion.  Nonetheless, it would be comforting to have a precise 
microscopic calculation which enabled us to obtain more quantitative confirmation of
the Matrix Theory rules.  

In particular, the leading order interaction (and of course the free string spectrum), 
automatically satisfy ten dimensional Lorentz invariance.  This was not an input, and
it is likely that the condition of Lorentz invariance completely fixes the light cone 
string perturbation expansion.  At second order in the string coupling, Lorentz 
invariance is achieved by a cancellation of divergences in the graphs coming from
iterating the lowest order three string vertex, with terms coming from higher
order contact interactions \ref\contact{M.B.Green, N.Seiberg,
Nucl.Phys. B299 (1988) 559; J.Greensite, F.R.Klinkhamer, Nucl.Phys. B304
(1988) 108, Nucl.Phys. B291 (1987) 557, Nucl.Phys. B281 (1987) 269. }.  Higher order contact terms correspond
precisely to higher dimension operators in the effective field theory expansion, and
are to be expected.  However, effective field theory arguments cannot determine
their coefficients.
It would be of the greatest interest to
have a microscopic demonstration of how this cancellation arises from the dynamics of 
SYM theory.  A successful calculation along these lines would, I believe, remove
all doubt about the validity and uniqueness of the matrix model.

The higher order contact terms raise another interesting question: does the
effective field theory expansion of SYM theory lead to an expansion of scattering
amplitudes in integer powers of $g_S$.   Many of the twist operators in the orbifold
conformal field theory have fractional dimensions.  Naively this would seem to lead to
fractional powers of $g_S$ in the effective Lagrangian.  However, as described above,
there is additional, hidden,  $g_S$ dependence coming from ultraviolet divergences
in the iteration of lower dimension operators, whose OPE contains 
these fractional dimension operators.  The ultraviolet cutoff is of course 
$\sim 1/g_S$.  Thus, in principle, fractional powers of $g_S$ might cancel in the
final answer.   A particular example of such a cancellation (in this case a cancellation
guaranteed by SUSY) can be seen in an old paper of Greensite and Klinkhammer \contact\ .
Again, it is frustrating not to have a general understanding of why such
cancellations occur.   

I would like to end this section by discussing the peculiar relationship of the
matrix model formalism to the string bit formalism of Thorn \thorn\ .  It is clear that
Matrix Theory does build strings out of bits - but each bit is an entire quantum
field theory.  This is another aspect of the fact that the compactified matrix theory
has more degrees of freedom than its infinite volume limit.  The partons of the
eleven dimensional theory are truly structureless, but those of the theory compactified
on a circle are two dimensional quantum fields. On the other hand, the process of taking
the large $N$ limit wipes out most of this structure.  The only states of the system with
energy of order $1/N$ are those in which the individual parton strings are unexcited and
only long \lq\lq slinkies \rq\rq with wavelengths of order $N$ are dynamical degrees
of freedom.  In the large $N$ limit which defines conventional string theory, the 
complex, compactified partons return to their role as simple string bits.

This discussion makes it clear that there are {\it two} processes of renormalization
going on in matrix string theory.  In the first, taking $g_S$ to be a small but finite
number we integrate out degrees of freedom with energies of order $1/g_S$ and obtain
an effective field theory for degrees of freedom whose energy scale is $g_S$ independent.
Then we take $N$ to infinity and obtain an effective field theory for degrees of 
freedom with energy of order $1/N$.  It is an accident of the high degree of SUSY
of the present system that the second renormalization step is rather trivial.  
SUSY guarantees that the system produced at the first stage is a conformal field theory.
As a consequence, the effective Lagrangian of the the long strings is identical
to that of the partonic strings.

Consider a hypothetical matrix model which leads in an analogous manner to perturbative
string theory on a background preserving only four spacetime SUSY charges.  There is then
no reason to expect the effective Lagrangian at the first stage of renormalization
to be a conformal field theory.  $g_S$ is a finite number and the scale of energies being
integrated out is finite.  Instead we expect to obtain, to power law accuracy in $g_S$,
a general renormalizable Lagrangian consistent with the symmetries.  Now take the large
$N$ limit.  The low energy degrees of freedom, with masses of order one
 will now renormalize the effective Lagrangian of the degrees of freedom with energies 
of order $1/N$.  Since $N$ is truly taken to infinity, the result must be a
conformal field theory.   Thus, unlike the case of maximal SUSY, we can expect to
obtain conventional looking string physics, and in particular the constraints on the
background coming from the vanishing of the beta function, only in the large $N$ limit.

\newsec{$F = M A_{T^2 \rightarrow 0}$}

The Aspinwall-Schwarz \ref\as{P.Aspinwall, Nucl.Phys.proc.Suppl. 46
(1996) 30, hep-th/9508154; J.H.Schwarz, Phys. Lett. B367 (1996) 97, hep-th/9510086.}\ equation which stands at the head of this section describes how
a general F theory compactification\ref\vafaf{C.Vafa, Nucl.Phys. B469
(1996) 403, C.Vafa, D.Morrison, Nucl.Phys. B473 (1996) 74,
Nucl.Phys. B476 (1996) 437.} emerges from M theory\foot{Aspinwall and
Schwarz described this for the \lq\lq father of all F theory compactifications \rq\rq -
ten dimensional IIB string theory.  The general rule was first enunciated by N. Seiberg, 
as recorded in footnote $1$ of the first reference of \vafaf\ .}.  M theory is compactified on an elliptically
fibered manifold $X$, and the area of the fibers is then scaled to zero.   This produces
a theory with $12 - d_X$ noncompact dimensions.  The simplest example is the Type IIB
string theory.   

In Matrix Theory, this is described in terms of the $SYM_{2+1}$ construction
discussed in the previous section.  We will see how to derive both IIA and IIB string
theory quite explicitly from this simple Lagrangian, understand both the duality
between them and the $SL(2,Z)$ symmetry of the IIB theory and derive the transverse $SO(8)$ 
invariance of the ten dimensional IIB theory in a nonperturbative manner.
We will also see quite clearly how it comes about that the IIB string is chiral.
 
Many of these results have been derived previously from considerations of duality and
SUSY.  I believe that the proper way to understand the relationship between these two
derivations is by analogy with chiral symmetry in QCD.  Many properties of the 
strong interactions can be understood entirely in terms of chiral symmetry.
The QCD Lagrangian on the other hand is, in principle, a tool for deriving {\it all}
the properties of the strong interactions.   It incorporates the symmetry principles
and enables us to fill in the dynamical details which are left blank in the 
chiral Lagrangian description of the strong interactions.   

Let us recall how IIA string theory on a circle derives from $SYM_{2+1}$.  One radius
$L_1$ of the M theory torus is taken much smaller than $\lp$ and the other much larger.
Correspondingly $\Sigma_1$ in the field theory torus is much larger than $\Sigma_2$.
We do a Kaluza-Klein reduction of the theory to fields which are functions only of
$\sigma_1$.  In this limit, the $1+1$ dimensional gauge field $A_1 (\sigma_1)$ has only
one dynamical degree of freedom, its Wilson loop.  The effective gauge 
theory is strongly coupled and the conjugate variable to the Wilson loop is frozen
in its lowest energy eigenstate.  The other component, $A_2$, of the gauge potential
becomes a compact scalar field which represents the string coordinate in the compact
direction.  

The IIB limit of the theory is defined by taking the area of the M theory torus to zero,
at fixed complex structure.  Using the dual relation between the SYM and M theory tori,
and the fact that the SYM coupling is given by $g_{SYM}^2 \propto
\Sigma_1 \Sigma_2$, we see that in this
limit we get a strongly coupled gauge theory in infinite volume.  In $2 + 1$ dimensions
, the SYM coupling is relevant and defines the mass scale for a set of confined gauge
invariant excitations.  We are taking the limit of infinite mass.  The limiting theory
will be a fixed point of the renormalization group, describing the massless excitations 
of the theory.  We will argue in a moment that such massless excitations definitely
exist, so the fixed point is not trivial.  It will be important to decide whether it is
an interacting fixed point or an orbifold of a free theory and we will see that several
arguments indicate the former.   In either case, the resulting theory is scale invariant.
Thus, although the volume is going to infinity, we can do a trivial rescaling of all
correlation functions in the theory to relate them to a theory on a torus of fixed volume, 
or with a nontrivial cycle of fixed length.  The possible backgrounds are then parametrized
by the complex structure of a two torus (which we will always take to be rectilinear
for simplicity) and there is an obvious $SL(2,Z)$ symmetry of the IIB physics
(spontaneously broken by the background).  Thus, as usual in an eleven dimensional
description, the strong weak coupling duality of IIB theory is manifest.

To show the existence of massless excitations, we go to the moduli space.  As usual,
this is characterized by the breaking of $U(N)$ to $U(N_1 ) \times \ldots \times U(N_k )$,
and parametrized by $k$ copies of a $U(1)$ gauge multiplet, with a permutation symmetry
relating those $U(1)$'s which have the same value of $N_i$.  The kinetic energy of
these fields (in terms of canonical variables) is $o(1/N_i )$.  In the strong coupling
infinite volume limit, it is convenient to rewrite the moduli space Lagrangian in terms of
dual variables.  With appropriate normalizations it takes the form
\eqn\IIBmodham{L = \sum_{i =1}^8 (\partial_{\mu} x^i )^2}
The variable $x^8$ is proportional to the dual photon field, $\phi$, defined by
\eqn\dualphoton{F_{\mu\nu} =
\epsilon_{\mu\nu\lambda}\partial_{\lambda}\phi .}
For a finite area torus, this variable is periodic.  Indeed, its canonical momentum
is the magnetic flux density, which is quantized in units of the inverse volume of the
SYM torus, or equivalently, the volume of the M theory torus.  In the IIB limit the
unit of quantization goes to zero and $x^8$ becomes an unbounded variable.

There is thus an $SO(8)$ symmetry of the moduli space Lagrangian which relates the
dual photon field to the seven scalars in the gauge multiplet.  This will turn out to be
the transverse rotation symmetry of uncompactified IIB string theory.  Indeed, the moduli
space contains a huge set of candidates for spacetime particle scattering states.  As in
the IIA theory, one can argue formally that in the $g_{SYM}^2 \rightarrow \infty$ limit
all finite energy states are describable on the moduli space, which is a $2 + 1$ dimensional
supersymmetric field theory containing $8N$ scalar fields living in the target space
${\bf R}^{8N} / S_N$.   Low energy excitations in the large $N$ limit are obtained by 
choosing fields in the twisted sector
\eqn\twisted{x_I (\sigma_1 + 2 n_1 \pi , \sigma_2 + 2 n_2 \pi ) = {(S_{(1)})^{n_1}}_I^J 
 {(S_{(2)})^{n_2}}_J^K x_K (\sigma_1 , \sigma_2 )}
with permutations $S_{(1,2)}$ of cycle length 
$ N^{\ha}$\foot{In order to give all matrix elements of
the diagonal matrix an effective long periodicity, we must write the
space as a tensor product choose the
permutations for orthogonal cycles to act on different factors of the
tensor product.  The lowest possible energies for two dimensional
configurations are of order $N^{\ha}$. We can get lower energy
configurations, which are one dimensional by choosing our fields to be
independent of one of the coordinates and to lie in a twisted sector
with cycle length $N$ with respect to the other.}.  
Morally speaking the number of excitations here is what one
would expect from a membrane.  We have, as yet, no argument that any of these excitations
are stable or metastable.  Following the discussion in eleven dimensions
 we could set up a path
integral to calculate the S matrix, but, in contrast to the gravitons of that discussion,
we do not have any indication that the single
particle states are stable.  In the IIA limit, metastability of the string states followed
from the fact that as $g_S$ went to zero, all string interactions vanished.  This in turn
followed from the nonrenormalization theorem for the orbifold Lagrangian.   

In $1 + 1$ dimensions, there is no superconformal algebra with $16$
SUSYs\ref\nahmnati{W.Nahm, Nucl.Phys. B135 (1978) 149; 
N.Seiberg, hep-th/9705117 .}.  
If one attempts to construct one by commuting SUSY generators with conformal generators, 
one finds an algebra which is not generated by a finite number of holomorphic and
antiholomorphic currents.  It is plausible that such an algebra can be a symmetry of
a conformal field theory only if it is free.  This is consistent with the fact that all
perturbations of the orbifold conformal field theory are irrelevant.  By contrast, in
$2 + 1$ dimensions, there is a superconformal algebra with $16$ SUSYs \nahmnati\ .
Thus, although the moduli space Lagrangian is not renormalized away from the orbifold
points, there may be a nontrivial conformal field theory describing dynamics at the
orbifold points.  Indeed, this {\it must} be the case if we believe that the matrix
model correctly describes Type IIB string theory.  In $1 + 1$ dimensions, the conformal
limit of the SYM theory coincided with the weak coupling limit of IIA string theory.
Here however, the conformal limit takes us to uncompactified IIB string theory at a 
value of the coupling given by the imaginary part of the $\tau$ parameter of the M
theory torus.  Thus, generically, this limit should describe a strongly interacting
theory.   This can only be true if a nontrivial conformal field theory resides at
the singular points of the orbifold\foot{For the simplest case, $U(2)$, this theory
also describes the infrared dynamics of two coincident Dirichlet 2 branes.}.

We know very little about this theory, which is a close cousin of the famous $(2,0)$ 
fixed point theory in $5 + 1$ dimensions.  However, we do know that it is exactly
$SO(8)$ invariant.  The superconformal algebra with $16$ SUSY generators includes
an $SO(8)$ R symmetry .  The SUSY generators transform as $Q_{A\alpha}$ where $A$ is a
spinor index of $SO(8)$ and $\alpha$ is a $2 + 1$ dimensional spinor index.  It
follows from the algebra that both SUSY generators are in {\it the same} spinor of
$SO(8)$.  When $SO(8)$ is interpreted as a spacetime symmetry, this tells us that
the theory is chiral in spacetime.  It is easy to see that the action of this
$SO(8)$ symmetry on the moduli space is precisely that of the explicit symmetry which
we have exhibited above \bs\ .
 The upshot of this discussion is that the
interacting IIB theory defined by the matrix model has an $SO(8)$ transverse
symmetry which rotates the momentum component defined as a limit of membrane
winding numbers into the $7$ noncompact components.  The SUSY algebra of the
resulting ten dimensional theory is the IIB algebra and the theory is chiral in
spacetime.   An alternative derivation of this rotation symmetry, which uses the
compactification of Matrix Theory on a three torus, was given by Sethi
and Susskind\ref\ss{S.Sethi, L.Susskind, Phys.Lett. B400 (1997) 265.}.

We can extract more information from the superconformal algebra if we make the
assumption, that the long distance, zero longitudinal
momentum transfer scattering of excitations of IIB theory can be encoded in
an effective Lagrangian on the moduli space.  The purely bosonic part of this action 
will be quartic in derivatives of the scalar fields $R^A$, which represent the differences
in transverse center of mass coordinates of the two objects we are scattering.  
From the discussion above, we know that the moduli space Lagrangian has spontaneously
broken superconformal invariance, where the VEVs of the $R^A$ are the symmetry breaking
order parameters.  Standard nonrenormalization theorems tell us that the kinetic term
of these fields is canonical, so that they have dimension $\ha$.  It then follows from
scale and $SO(8)$ symmetries that the quartic term in the Lagrangian has the form
\ref\bfssIIB{T.Banks, W.Fischler, N.Seiberg, L.Susskind, hep-th/9705190 .}
\eqn\quartic{A {v^4 \over r^6} + B {({\bf rv})^2 v^2 \over r^8} + C
{({\bf rv})^4 \over r^{10}}}
The coefficients $A,B,C$ are undetermined by this argument, though I suspect that the
full superconformal algebra determines at least their relative sizes.

One can also approach this calculation using instanton methods in the $2+1$ dimensional
gauge theory.  Since we are interested in the strong coupling limit one must hypothesize
that there is some sort of nonrenormalization theorem for the quartic operator
which tells us that the instanton calculation is exact.  Unfortunately the multiinstanton
calculations of \ref\dkm{N.Dorey, V.Khoze, M.Mattis, hep-th/9704197 .} do not reproduce the correct $SO(8)$ invariant behavior.
Either the hypothesis made in \dkm\ about multiinstanton moduli space is incorrect, or
the hypothesis of a nonrenormalization theorem must be modified to a claim that instantons
plus a finite set of perturbative corrections to them give the exact answer (for this
is the nature of the correct answer).  Such
theorems are not unheard of \ref\edthm{E.Witten, Comm. Math. Phys. 141,
(1991), 153; J. Geom. Phys. 9, (1992), 303.}.  
If we Fourier transform the answer to
give the amplitude for a fixed value of momentum in the \lq\lq membrane winding \rq\rq
direction then instanton methods seem to give better results.

If we compactify the IIB theory (which is the same as going to a finite volume M theory
torus) then the leading effect on this Fourier transformed amplitude is to quantize the
allowed values of the momentum.  A fixed value of momentum corresponds to a fixed instanton
number.  In particular, for the single instanton amplitude, where there is no
ambiguity about instanton moduli space, we can use the result of Pouliot and Polchinski
\pp\ that already came into our discussion of membrane scattering with longitudinal
momentum transfer.  Strictly speaking, one should , in the present context, do the
instanton calculation at finite volume.  However, we are really interested in the
limiting form of the amplitude in decompactified IIB theory, and the calculation of \pp\ 
should be sufficient for this.   This calculation tells us that $B = C = 0$ and fixes the
value of $A$.   The result is in agreement with the expectations of IIB supergravity
in ten dimensions.  Note that our derivation made no use of string perturbation theory.

We will end this section by indicating how IIB perturbation theory can be derived
from matrix theory.  Following Aspinwall and Schwarz \as\ , one takes the limit of a torus 
with two sides of very different size.  Along the moduli space we can do a Kaluza-Klein
reduction of the $2+1$ dimensional theory.  The low energy degrees of freedom are
functions of a variable parametrizing the long side of the gauge theory torus.  Following
precisely the logic of our IIA discussion we find the Fock space of IIB Green-Schwarz
string theory.  In this $1+1$ dimensional reduction of the theory there are again
nonrenormalization theorems which guarantee that the leading correction to the orbifold
Lagrangian is of dimension three and has precisely the form of the conventional
three string vertex in light cone gauge.   Furthermore, in units of the perturbative 
IIB string tension, the degrees of freedom with nonzero Kaluza-Klein momentum along the
short direction of the gauge theory torus, have masses of order the ratio ${\Sigma_1 \over
\Sigma_2}$ of the long to short sides of the torus.   Thus the IIB string coupling
is of order ${L_1 \over L_2}$, the ratio of short to long sides of the M theory torus.
We cannot verify that the correct numerical coefficient is obtained without a microscopic
derivation of the effective field theory from the nontrivial fixed point theory in
$2 + 1$ dimensions.  

Matrix Theory also provides a vivid description of the D strings of IIB theory.  For
example, $(0,1)$ D strings are simply fields which depend only on the \lq\lq 
short \rq\rq coordinate of the SYM torus.  Their tension is of course reproduced
exactly since it is given by a BPS formula.  These states are not in general stable, 
except for the infinite D string.

T duality of weakly coupled string theory follows simply
from the $2+1$ dimensional dual transformation discussed above.  Both the weakly coupled
compactified IIA theory and the weakly coupled compactified IIB theory correspond
to SYM theory on a torus with finite volume and sides of very different length.
The compact IIA coordinate is the Wilson line around the short direction of the SYM
torus, which can be thought of as a scalar field depending on
 the long coordinate.   The compact IIB coordinate
is the scalar dual to the photon field, which in the Kaluza-Klein reduced theory is
a function of the same coordinate.  The $2+1$ dimensional duality relation reduces
to $1 + 1$ dimensional duality between these scalars.

Finally, we note a puzzle connected with the IIB theory which we have
constructed.  Our calculation can be done on a two torus with non
orthogonal sides.  In perturbative IIB string theory, the real part of
the modulus of the torus is the expectation value of a Ramond-Ramond
scalar, and does not show up in any perturbative amplitude.  The leading
weak coupling contribution to $Re \tau$ dependent amplitudes comes from
D instantons.  This is far from obvious in Matrix Theory.  In
particular, for finite $N$, modes of the two dimensional fields which
depend on both coordinates of the torus are sensitive to $Re \tau$.
When we make the Kaluza-Klein expansion which determines the Type IIB
perturbation series, we will find $1 + 1$ dimensional fields with $Re
\tau$ dependent masses of order $1/g_s$.  Naively, these should show
up as finite order, $Re \tau $ dependent contributions to the
perturbation expansion.   It seems that these contributions could at
best vanish in the large $N$ limit, when most of the microscopic modes
of the matrix field theory are frozen out.  There seems to be little
chance that the finite $N$ small $g_s$ expansion could be independent of
$Re \tau$.  We will return to this point when we discuss Discrete Light
Cone Quantization. 
\newsec{Compactification on a Three Torus}

M theory compactified on a three torus has an $SL(2,Z) \times SL(3,Z)$ duality
group.  $SL(3,Z)$ is the obvious group of discrete diffeomorphisms of the three torus.
If we choose a torus with one small cycle, to obtain $T^2$ compactified IIA string
theory, then the diagonal $SL(2,Z)$ is the T duality group of this string theory.

It was shown in the beautiful papers \grt\ ,
\ref\ld{L.Susskind, hep-th/9611164.} that the the full M
theory duality group is simply understood in Matrix Theory as the product of the
geometrical $SL(3,Z)$ of the SYM torus, with the Olive-Montonen $SL(2,Z)$ 
duality symmetry of $N = 4$ , $D = 4$ SYM theory.   Using these symmetries and
the conformal invariance of $SYM_{3+1}$ it is easy to show that all limits of the
moduli space correspond either to weakly coupled IIA string theory or to decompactified
M theory.   

\newsec{Four and More}

It is in four compact dimensions that the true nature of M theory begins to show itself.
The SYM prescription for compactification obviously runs into trouble at this point,
because the SYM theory is nonrenormalizable.  As long ago as December of 1996, N. Seiberg
suggested in discussions at Rutgers that one way to define the $4 + 1$ dimensional
SYM theory was via compactification of the $5 + 1$ dimensional fixed point theory
with $(0,2)$ SUSY.  At the time this seemed a rather arbitrary prescription.  In general
there are many ways to define a nonrenormalizable theory as a limit of more complex
high energy physics.  The key reason that the $(0,2)$ prescription is the right one
was pointed out by M. Rozali \ref\rozali{M.Rozali, hep-th/9702136.}: it reproduces the $SL(5,Z)$ duality 
group which one expects for M theory compactified on a  4 torus.  Rozali argued that
the $(0,2)$ prescription could be derived from the dynamics of SYM by considering
threshold bound states of instantons.  This I believe to be incorrect.  The semiclassical
treatment of the threshold bound state problem immediately runs into all of the
problems of nonrenormalizable field theory.  The short distance instanton interaction
cannot be calculated correctly in SYM theory.  I believe that it is also unlikely
that any conventional regulated version of SYM theory will reproduce the correct
dynamics.  When the $(2,0)$ field  theory is reduced to SYM theory by compactification, the
\lq\lq n instanton states \rq\rq are just the lowest energy states carrying Kaluza-
Klein momentum.  They are created by local fields in $4 + 1$ dimensions, with a 
short distance cutoff of
order the  SYM coupling.  Thus, for arbitrary $n$, these states have a size of order
$g$.  On the other hand, in cutoff field theory one expects the size of an $n$ particle
bound state to grow with $n$.  All this being said, there is currently a proposal
under study \ref\aharony{O.Aharony, M.Berkooz, S.Kachru, N.Seiberg,
E.Silverstein, hep-th/9707079.} for deriving the $(0,2)$ theory from a model of instanton
bound states.  In this case, the SYM dynamics itself is derived from the instantons.
We will discuss this further below.  Berkooz, Rozali and Seiberg
\ref\brs{M.Berkooz, M.Rozali, N.Seiberg, hep-th/9704089.} put together
the two ideas about the relation between SYM and the $(0,2)$ theory, and suggested that
the generalization to the five torus would be the compactification of 
a Poincare invariant theory without gravity, which was not a quantum field theory.  
Seiberg \ref\ns{N.Seiberg, hep-th/9705117.}
then showed that the existence of such a theory was guaranteed by known facts about
duality and M theory.

In this section I will attempt to present these developments (which all postdate the
Trieste lectures on which this text is based) in a logical manner and to
provide some clues to the general structure which lies behind them.  In giving lectures
about Matrix Theory over the past year, I have constantly had to remind the
audience that the theory was in flux and that some of the ideas which I was
describing are likely to turn out to be wrong in large or small ways.  The time has
come to again draw such a demarcation line.  Up until this point I am reasonably
confident that everything which I am relating to you will stand the test of time.
This is not true for the material in the following subsections and for ideas in later
sections which depend on it.

\subsec{Longitudinal Charges}

The material in this subsection is drawn from \bss\ ,
\bs\ , and \ns\ .
In our discussion of the normalization of parameters in SYM theory, we demonstrated
that electromagnetic fluxes in the SYM theory appeared as central charges in the 
SUSY algebra of the uncompactified dimensions.  These were Kaluza-Klein momenta in
spacetime (electric fluxes) and transversely wrapped two brane charge (magnetic fluxes).
We also pointed out that there were charges corresponding to longitudinally wrapped
membranes, and fivebranes wrapped around the longitudinal and four transverse
directions.  

We would now like to emphasize the fact that the longitudinal membrane charges are
nothing but the momenta of the SYM field theory.  A quick way to see this is to remember the derivation of the SYM prescription
for compactification from the dynamics of D0 branes in IIA string theory.  There,
the SYM momenta are interpreted as the winding numbers of strings between zerobranes.
But IIA strings are longitudinally wrapped membranes, so SYM momenta are just longitudinal
membrane charges.  In low dimensions, the duality group of M theory transforms
five brane charges into membrane charges.  Thus, one would expect to have operators
carrying all such charges.  The reason that the five brane charges are unnecessary
in compactification with less than three compact transverse 
dimensions is that {\it on a torus
with four or fewer dimensions (including the compactified longitudinal direction)
there are no finite energy wrapped fivebrane states.}  This leads to the conjecture that
in general, the IMF Hamiltonian of M theory will contain canonical degrees of freedom
carrying the quantum numbers of each type of finite energy wrapped
longitudinal brane.

In particular, on a transverse four torus, in addition to the four dual momenta (
corresponding to the four types of wrapped longitudinal membrane) we should
expect to have a fifth charge corresponding to longitudinal five branes.  In a limit
of moduli space in which all states carrying the fifth charge are heavy, the theory
should reduce to $SYM_{4+1}$.  Adding the requirement that the theory have $16$ super
charges uniquely specifies the $(0,2)$ fixed point theory.  This flows to $SYM_{4+1}$
after compactification of one dimension.  The fixed point theory is scale invariant, so
the radius of the smallest circle on the five torus is conveniently chosen
to define units.  It defines
the SYM coupling, and through it the eleven dimensional Planck scale.

This definition is manifestly covariant under the $SL(5,Z)$ diffeomorphism 
group of the five torus, which is the duality group of M theory on the four torus.
Furthermore, the $(0,2)$ theory contains self dual antisymmetric tensor fields.
The electric (which are the same as the magnetic ) fluxes of this field transform
properly as a second rank antisymmetric tensor of $SL(5,Z)$.  Under Kaluza-Klein reduction 
on one circle, this splits up into a vector (interpreted as spacetime momenta in the
four compact directions) and an antisymmetric tensor (interpreted as transverse
membrane winding number).  Note that when all radii of the 5 torus are comparable, so
that it is not sensible to make a Kaluza-Klein reduction of degrees of freedom, there is
no natural definition of the compact directions of space.  This is analogous to
the situation in $T^2$ compactification when both radii are of order the Planck scale.
There also there are two possible definitions of spacetime, one appropriate to the
eleven dimensional limit and the other to the IIB limit.  In the Planck scale regime
neither definition is appropriate.   Of course, there is a hint of this phenomenon 
which goes back to the first papers on T duality.   The matrix model makes it abundantly
clear that space is a derived concept in M theory, one whose utility depends on 
a Born-Oppenheimer separation of degrees of freedom which is not always valid.

Perhaps it is worthwhile to pause here to give a description of what is known about
the $(0,2)$ superconformal field theory.  The M theory fivebrane has a selfdual
antisymmetric tensor gauge multiplet of $(0,2)$, $5+1$ dimensional SUSY propagating
on its world volume.  It contains a two form potential with self dual field strength, five
scalars whose zero modes describe the transverse position of the fivebrane in
eleven dimensional spacetime, and sixteen real fermion fields transforming as two
complex chiral spinors of the six dimensional Lorentz group.  The fivebrane is also
a D-brane for M theory membranes \ref\andypaul{A.Strominger,
Phys.Lett. B383 (1996) 44, hep-th/9512059; P.K.Townsend, Phys.Lett. B373
(1996) 68, hep-th/9512062.}.   Now if we consider $n$ parallel
fivebranes, there are $o(n^2)$ strings on the individual fivebrane world volumes,
whose tension goes to zero in the limit that the five branes are all in the same
transverse position.  This limit then would seem to describe a $5+1$ dimensional
dynamics which decouples from gravity.  It is the most supersymmetric of what have come
to be called {\it tensionless string theories}.
Seiberg\ref\ns?{Phys.Lett. B390 (1997) 169, hep-th/9609161.}, and
Witten\ref\ed?{E.Witten, Mod. Phys. Lett. A11, (1996), 2649, hep-th/9609159.}
have argued that these are in fact superconformal field theories at nontrivial fixed
points of the renormalization group.  The general $(0,2)$ theory has a moduli space
consisting of $r$ copies of the tensor multiplet, where $r$ is the rank of some self
dual Lie group\ref\natisixteen{N.Seiberg, hep-th/9705117}.  
The electric and magnetic fluxes on the moduli space allow us
to define a set of charges.  Dirac-Nepomechie-Teitelboim quantization arguments
tell us that these charges can lie in the weight lattice of a self dual group.
The group appropriate for $n$ fivebranes, and also for the matrix model, is $U(n)$.

The practical utility of the $(0,2)$ prescription will depend on our ability to
formulate and solve it.  I will report below on a recent proposal for doing so, but
we are as yet far from the goal.   First however, I want to briefly describe the
prospects for further compactification.  Following our rule of counting finite
energy longitudinal branes, we should expect degrees of freedom labelled by
ten integers $(P_L , P_R )$ (each letter stands for a five vector) transforming in the
ten dimensional representation of the $O(5,5)$ duality group.  These correspond to the
5 ways of wrapping longitudinal membranes and 5 ways of wrapping longitudinal 5 branes
around a five torus.   Furthermore, the moduli space of the theory is
$O(5,5,Z)\backslash O(5,5)/[O(5)\times O(5) ]$.  There are limits in moduli space for which one 
of two sets of five linear combinations $P_L \pm P_R$ become continuous (in the sense that
the cost in energy for each integer unit goes to zero).  In these limits we expect to
get either $5+1$ dimensional SYM theory (the limit in which all five M theory radii get
large) or the $5+1$ dimensional $(0,2)$ theory (the limit in which one of the five M theory
radii is infinite).  The duality group here is identical to the T duality group of
a string theory compactified on a five torus, as is the moduli space. 
The two low energy limits with five continuous momenta are T dual to each other
in this sense (T duality in one circle).  This led the
authors of \brs\ to argue that the relevant theory here could not be a local 
quantum field theory on a five torus.    A quantum field theory has a unique local
stress energy tensor which represents the response to infinitesimal changes in the metric.
Near the fixed points of the T duality group, there are infinitesimal changes in the metric
which do not change the theory.  This is argued to lead to unacceptable behavior
for the Green's functions of the stress tensor.  They argued that this was evidence
for a Poincare invariant theory without gravity which was not a local field theory.
They further suggested that it was some sort of noncritical string theory.

In \ns\ Seiberg identified another context in M theory in which a model with precisely
these characteristics arose (in the same way that the $(0,2)$ fixed point theory arises
both in the matrix model and in the theory of $n$ coincident M theory fivebranes in
uncompactified eleven dimensional spacetime).   Consider $n$ Neveu Schwarz fivebranes
in either Type IIA or Type IIB string theory in the limit in which the string coupling
goes to zero.  Begin with the IIB case, where we can relate NS fivebranes to
D fivebranes by S duality.  When the D fivebranes are close together, the low energy
theory is a $5+1$ dimensional SYM theory with coupling $g_D^2 = g_S l_S^2 $.  By S duality, 
coincident NS 5 branes should carry a SYM theory with coupling $g_{NSB}^2 = ({1\over g_S})  
({g_S l_S^s}) = l_S^2$.  Thus, the interactions between NS fivebranes are finite
even when the string coupling goes to zero\foot{In the semiclassical picture of the
NS fivebrane\ref\chs{C.G.Callan, J.A.Harvey, A.Strominger, in
Proceedings of the 1991 Spring School on String Theory and Quantum
Gravity, Trieste, Italy, hep-th/9112030.} 
this can be attributed to the fact that the dilaton varies in 
space and the coupling goes to infinity in an infinite tube at the core of the
fivebrane.}.  On the other hand, in this limit the fivebranes decouple from string states 
which can propagate in the bulk, and in particular, from gravity.  Thus, we expect
that we are left with a Poincare invariant theory without gravity.

Now consider the same theory compactified on a five torus, with the fivebranes wrapped
on the torus.  The subgroup of the U duality
group of string theory which leaves $g_S = 0$ invariant is precisely the T duality
group described above.  Thus our compactified Poincare invariant theory is invariant
under T duality.  In particular, upon inversion in a single circle, it becomes the theory 
of IIA NS fivebranes with vanishing string coupling.  Since these are just 
reduced M theory fivebranes, there is a low energy limit which is just the $(0,2)$ theory.
Thus, the theory of compactified Type II fivebranes with vanishing string coupling
has all the characteristics necessary for the construction of a matrix model of M
theory compactified on a five torus.

Seiberg's most compelling evidence for this identification is his demonstration that
the zero coupling limit of the fivebrane theory contains states with the characteristics 
of all the wrapped two brane and fivebrane states of M theory on a five torus.  
To make this identification one must first choose a dictionary for translating the radii
and string tension which parametrize the NS fivebrane theory into M theory parameters.
There are enough states in the theory to guarantee nontrivial checks of the formulae
even after the identification is made.
In 
particular the wrapped transverse fivebrane is a bound state of a D5 brane and the $n$
IIB NS fivebranes.

Recently a unified description of the various toroidal compactifications
of Matrix Theory has been provided by Sen\ref\sen{A.Sen,
hep-th/9709220. } and Seiberg
\natiproof\ \foot{This prescription was independently invented by
L. Susskind, and described to the author in July of 1997.}.  Seiberg has
used it as the basis of a {\it proof} that the matrix prescription is in
fact the Discrete Light Cone Quantization (DLCQ) of M theory.

The proof is easy to describe.  DLCQ is compactification on a lightlike
circle, defined via periodic identification along the vector $2\pi (R,R, {\bf
0_9})$ .  This is the limit of compactification on a spacelike circle
defined by periodic identification along $(R, R + R_s ,{\bf 0})$., as
the Minkowski norm of the compactification direction, $- R_s^2 $, 
 is taken to zero.
If we assume that compactified M theory has a vacuum state invariant
under Lorentz transformations in the uncompactified directions, then
this is equivalent to compactification on a spacelike circle $(0, R_s,
{\bf 0_9})$.   But M theory on a spacelike circle of zero radius is
free Type IIA string theory.

In DLCQ, we are instructed to study the light cone Hamiltonian (energy
minus longitudinal momentum) in a sector of fixed positive longitudinal
momentum.  Using Lorentz invariance, this translates into a sector with
fixed positive momentum around the spacelike circle.  In Type IIA
language, we are instructed to work in a sector with fixed D0 brane
number $N$ and to subtract the zerobrane mass $N/R$ from the
Hamiltonian.  We then obtain a spectrum of zero brane kinetic energies
which go to zero for fixed transverse momentum (of order the eleven
dimensional Planck scale), as $R_s$ goes to zero.  Seiberg shows that
this is the scale of energies which are finite in the original, almost
lightlike, frame.   We should thus try to write down the Effective
Hamiltonian for all states which have energies of this order or lower.  
The work of \kpdfs\ , and \dkps\ tells us that we should then include the SYM
interactions between D0 branes, which have the same scaling as their
kinetic energy.

In the compactified theory, with compactification radii of order the
eleven dimensional Planck scale (and thus small in string units), the
cleanest way to isolate the relevant degrees of freedom and interactions
is to do a T duality transformation.  On tori of dimension less than or
equal to three, this reproduces the SYM prescription for
compactification.  On the four torus, the T dual theory is the theory of
$N$ D4 branes, but the T dual IIA coupling is going to infinity.  Thus,
the proper way to view this system is as a set of $N$ fivebranes in M
theory, wrapped around the large \lq\lq eleventh \rq\rq dimension, and
the T dual four torus.  The scale of the (original picture) D0 brane
kinetic energies is the same as that of the tensionless strings on
the T dual M theory fivebranes.  Thus the effective theory is the
$(0,2)$ conformal field theory, with structure group $U(N)$.

On the five torus the T dual theory is that of $N$ D5 branes in strongly
coupled IIB string theory.  By S duality, this is the same as the system
of IIB Neveu-Schwarz fivebranes at infinitely weak coupling and fixed
string tension, $M_S^2$, which is
the proposal of Seiberg described above. We note however, that Maldacena
and Strominger \ref\maldstrom{J.M.Maldacena, A.Strominger,
hep-th/9710014.} have recently argued that this system
also contains a new continuum of states above a gap of order $M_S$.
These are abstracted from a SUGRA description of the excitations of the
NS 5 brane system as near extremal black holes.  The string frame
description of these geometries contains an infinite tube which
decouples from the bulk as the string coupling goes to zero.
If the deviation from
extremality is taken to zero along with the string coupling, it is
argued that the coupling between the continuum of modes running up and
down the tube, and the excitations on the five branes, remains finite in
the limit.  It goes to zero only when $N$ is taken to infinity.

We will see that this continuum represents extremely bizarre physics
from the Matrix Theory point of view, but we reserve that discussion
until we have described the even more bizarre situation on the six
torus.  Here, the T dual theory is that of D6 branes in strongly coupled
Type IIA string theory.  Again, the appropriate description is in terms
of eleven dimensional SUGRA.  That is, the theory consists of M theory
compactified on a circle with radius $R_T \rightarrow \infty$, with $N$
Kaluza-Klein monopoles (the original D0 branes) 
wrapped around a six torus whose size is the
T dual eleven dimensional Planck scale, 
$L_p$,(this is T dual to the original six
torus whose size is the original eleven dimensional Planck scale before
T duality).   In the limit we get the theory of $A_{N-1}$ singularities
interacting with SUGRA and wrapped around a Planck scale six torus.

Certain aspects of the physics are best understood before taking the
limit.  Then the Kaluza-Klein monopoles can be viewed as particles in
the uncompactified spacetime.  They have mass $R_T^2 L_p^{-3}$.  If the
T dual D0 branes are given finite momenta in the original Planck units,
then the KK monopoles must be given momenta $R_T L_p^{-2}$.  Their energy
is then finite in $L_p$ units.  But the situation regarding momenta is
strange.  The KK monopoles carry infinitely more uncompactified momentum
in the $R_T \rightarrow \infty$ limit than the supergravitons of
comparable energy.  Thus in the limit, KK monopole momentum is
conserved, while supergraviton momentum is not.   Indeed, this is the
only way that the relativistic supergraviton dispersion relation could
have been made compatible with the Galilean invariance which we require
for the light cone interpretation of the matrix model.   

From the point of view of the original M theory which we are trying to
model, the physics of this system is completely bizarre.  It says that M
theory (or at least DLCQ M theory) on a six torus 
contains a continuum of excitations
in addition to that described by the asymptotic multiparticle states in
ordinary spacetime.  These states carry finite light cone energy, but no
transverse or longitudinal momentum.   Scattering of M theory particles
can create these states and the energy lost to them need never appear in
the asymptotic region of M theory.  The asymptotic states (in the usual
sense) of M theory are not complete, and their S matrix is not unitary.
The situation is analogous to a
hypothetical theory of black hole remnants, except that the remnants are
zero momentum objects which fill all of transverse and longitudinal
space in the light cone frame.
Furthermore, it is clear that such a description is not Lorentz
covariant under the lightplane rotating transformations which we hope to
recover in the large $N$ limit.  The excitations described on the five
torus by Maldacena and Strominger produce a very similar picture, the
major difference being that their excitations are separated from the
spacetime continuum by a finite gap, and are therefore invisible at
sufficiently low energies.

There is reason to believe that both of these problems go away in the
large $N$ limit (and I think they must if the theory is to be Lorentz
invariant ).  KK monopoles repel supergravitons carrying nonzero
momentum (the only ones which couple in the $R_T \rightarrow \infty$
limit) around the KK circle, because the size of the circle goes to
zero at the center of the monopole.  In the $R_T \rightarrow \infty$  
limit, the circle has infinite radius everywhere apart from the position
of the singularity.   In the large $N$ limit, this repulsion becomes
infinitely strong.  Supergravitons localized at any finite distance from
an $A_{\infty -1}$ singularity have infinite energy\foot{This is a
colloquial description of the true situation, which is properly
described in terms of the scattering amplitude of the gravitons on the singularity.}.  Thus it is
plausible, though not proven, that they decouple in this limit.  Similar
remarks may be made on the five torus, where Maldacena and Strominger
have argued that the excitations propagating in the throat of the near
extremal black hole decouple because the Hawking radiation rate vanishes
in the large $N$ limit.  We will expand further on these remarks when we
discuss DLCQ below.

The problems encountered on the five and six tori may, in a way which I
do not yet understand, be precursors of a more evident problem in lower
dimensional compactifications.  DLCQ of a theory with four noncompact
dimensions is effectively a $2+1$ dimensional theory, and gravity
compactified to $2+1$ dimensions has infrared divergences if we require
the geometry to be static.  This is the origin of the
claim\ref\banksuss{T.Banks, L.Susskind, Phys.Rev. D54 (1996) 1677,
hep-th/9511193.}
that low dimensional string theories with static geometries 
do not have many states.  Zeroth order string perturbation theory misses
this effect and leads one to expect toroidal compactifications of any
dimension.  In fact, below three noncompact space dimensions (which
means three noncompact transverse dimensions in the case of DLCQ), we
should be studying cosmology. 

It should not have to be be emphasized that this line of reasoning is somewhat conjectural.
Several recent papers have claimed to construct a matrix model of M theory on a
six torus \ref\mM{A.Losev, G.Moore, S.Shatashvili, hep-th/9707250;
I.Brunner, A.Karch,hep-th/9707\break 259; A.Hanany, G.Lifschytz, hep-th/9708037
.} and there have even been some conjectures about complete
transverse compactification\ref\emilpaul{E.Martinec, hep-th/9706194;
P.K.Townsend, Talk given at Strings 97 Meeting, Amsterdam, June 1997,
hep-th/9708034.}.  The authors of \ref\egkr{S.Elitzur, A.Giveon,
D.Kutasov, E.Rabinovici, hep-th/9707217.} have
shown that much of the algebraic structure of the $E_d$ duality group
of the $d$ torus can be derived already from the prescription that the 
compactified matrix theory be invariant under the manifest symmetries of
the torus, combined with a consistent generalization of the electric
magnetic duality of the $3+1$ dimensional theory.
This whole subject is in a state of flux, and it is too early to tell what the
outcome will be.

Returning now to firmer ground, we will discuss Seiberg's construction of the
transverse fivebrane \ns\ .  Matrix Theory on $T^5$ is the $g_s \rightarrow 0$
limit of the theory of $N$ NS fivebranes wrapped on the five torus in Type II
string theory.  This theory is characterized by a single dimensionful parameter
$l_s$, and the geometric (and background 3 form field) data of the torus.
In this analog model, the transverse fivebrane is a bound state of the D5 brane
of IIB string theory with the NS fivebranes.  Its tension is of order $l_s^{-6}$.
In the IIB picture of the dynamics of the analog model, the low energy limit is
described by $5+1$ SYM theory.  Wrapped transverse membranes, and longitudinal
membranes and fivebranes are identified with various classical
configurations in this field theory. This is possible, even though some of the
relevant configurations are classically singular\foot{{\it e.g.} 
a single longitudinal fivebrane is a single instanton on a torus.}
and their quantum dynamics is surely
singular, because the SYM objects are localized and have long range fields.
Thus, although SYM is not a complete description of the physics at all energy
scales, one can get a picture of the relevant states.  This is similar to
the description of a grand unified 't Hooft-Polyakov monopole as a Dirac monopole
in QED.  On the other hand, the wrapped transverse fivebrane is uniform on the
SYM torus.  It is a state with constant energy density, rather than a localized
configuration.  Thus, it cannot be seen in the low energy SYM  theory.

To conclude this section, we will briefly summarize recent work
\aharony\ , \ref\ed{E.Witten, hep-th/9707093.}, 
which attempts to actually construct the Hamiltonians for the $(0,2)$ field theory
and noncritical string theory which describe Matrix Theory on $T^4$ and $T^5$ 
respectively.  For the moment, these constructions are restricted to the limit in
which the ``base space'' on which these theories live is noncompact six dimensional
Minkowski space.  Since the $(0,2)$ theory is a local field theory it should not
be too difficult to compactify it.  In the case of the noncritical string theory,
compactification may involve further conceptual problems.

The basic idea of \aharony\ and \ed\ is that the two desired theories are limits of
certain situations in M theory.  Since the matrix model gives us a definition of M
theory in some cases, we can try to use it to construct these theories.  This may 
seem somewhat circular, since we plan to use these theories to construct
Matrix Theory!
The point is that the uncompactified version of {\it e.g.} the $(0,2)$ theory is obtained
by studying the low energy dynamics of closely spaced fivebranes in {\it uncompactified}
 M theory.
If we choose a light cone frame, and orient the fivebranes so that the longitudinal
direction lies within them, then Berkooz and Douglas \ref\bd{M.Berkooz,
M.R.Douglas, Phys.Lett. B395 (1997) 196, hep-th/9610236.} have given us a complete
prescription for this system in Matrix Theory.  It is the theory of the low energy
interactions of
$k$ D4 branes and $N$ D0 branes in IIA string theory. This is the dimensional reduction
to $0 + 1$
dimensions of a six dimensional U(N) gauge theory with eight real
supercharges.  In addition to the Yang Mills vector multiplet, we have a hypermultiplet
in the adjoint representation and $k$ hypermultiplets in the fundamental.  In the quantum
mechanical reduction, the $5$ spatial components of the vector fields and the $4$ real
components of the adjoint hypermultiplet represent the $9$ (nonabelian) transverse
coordinates of excitations in eleven dimensional spacetime.  The vector components
are directions perpendicular to the fivebranes, while the adjoint components 
are transverse light cone directions which are in the branes.
There is a $U(k)$ global symmetry which acts on the fundamental hypermultiplets.
The weight lattice of this group is the charge lattice of
the $(0,2)$ theory we are trying to construct.  In IIA string theory we would have
a $4+1$ dimensional $U(k)$ SYM theory describing the self interactions of the $D4$ branes, but 
in Matrix Theory these degrees of freedom are dropped because they do not carry
longitudinal momentum.  The physics associated with these gauge interactions should
reappear automatically in the $N \rightarrow \infty$ limit.

The Berkooz-Douglas model describes longitudinal fivebranes in interaction with
the full content of M theory.  The Coulomb branch of the space of fields, 
where the components of the vector multiplet are large (but commuting so that the
energy is low), represents propagation away from the fivebranes, while the Higgs
branch, on which hypermultiplet components are large (but satisfy the D flatness
condition so that the energy is low) represents propagation within the fivebranes.
Mathematically, the Higgs branch is the moduli space of $N$ $SU(k)$ instantons.
We would like to take a limit in which the Higgs and Coulomb branches decouple
from each other.   
Viewed as a dimensionally reduced gauge theory, our model has only one parameter, 
the gauge coupling.  The Coulomb and Higgs branches describe subsets of zero
frequency (the quantum mechanical analog of zero mass) degrees of freedom which 
interact with each other via the agency of finite frequency degrees of freedom.
The gauge coupling is relevant, and if we take it to infinity all finite frequency
degrees of freedom go off to infinite frequency.  Thus the Coulomb and Higgs
branches of the theory should decouple in this limit.  The theory on the fivebranes, which
we expect to be the $(0,2)$ field theory, is thus argued to be the
infinite coupling limit of quantum mechanics on the Higgs branch.  

In other words, the claim is that the light cone 
Hamiltonian of $(0,2)$ superconformal field theory with
$U(k)$ charge lattice, is the large $N$ limit of the supersymmetric 
quantum mechanics on the moduli space of $N$ $SU(k)$ instantons.
The latter form of the assertion invokes a nonrenormalization theorem.
The instanton moduli space is Hyperkahler.  Presumably (but I do not know a precise
argument in quantum mechanics) the only relevant supersymmetric lagrangians for these 
fields are just free propagation on some Hyperkahler geometry.  The metric on the
moduli space is determined by the Hyperkahler quotient construction in the limit of weak
SYM coupling (that is, it is determined by plugging the solution of the D flatness
condition into the classical Lagrangian).  Furthermore, the SYM coupling can be
considered to be a component of a vector superfield.  Therefore the metric of the
hypermultiplets cannot be deformed and takes the same value when the coupling is
infinitely strong as it does when it is infinitely weak.

A stronger version of the assertion, which may be more amenable to checks, is that the
finite $N$ instanton quantum mechanics is the DLCQ of the $(0,2)$ field theory.

A similar set of arguments can be made for the the theory of \ns\ , described as the
weak coupling limit of $k$ NS fivebranes in IIA string theory \aharony\ , \ed\ .  We 
replay the above analysis, for fivebranes in
 M theory compactified on a circle of small radius, taking the
fivebranes to be longitudinal, and orthogonal to the circle.  This leads to a matrix
string theory which is just the dimensional reduction of the same six dimensional
gauge theory to $1 + 1$ dimensions\ref\dvvedaha{R.Dijkgraaf, E.Verlinde,
H.Verlinde, hep-th/9704018, 9709107;
 O.Aharony, M.Berk-ooz, S.Kachru, N.Seiberg,
E.Silverstein, hep-th/9707079; E.Witten, hep-th/9707093.}.
The $g_s \rightarrow 0$ limit of \ns\ is
again the strong coupling limit of the gauge theory, and the dynamics on the fivebrane
is the two dimensional conformal field theory of the Higgs branch.
In this context there has been some confusion about the appropriate Lagrangian
describing the conformal field theory and the reader is referred to the literature
for more details.  Even when this is sorted out, we will still be faced with the 
problem of compactifying this nonlocal, noncritical string theory.

Although the arguments supporting this \lq\lq matrix model for matrix models \rq\rq
approach are quite beautiful and convincing, I would like to point out a 
possible loophole, and some evidence that perhaps this construction fails.
While the general logic of this construction is impeccable, there is one point at which
error could creep in.  In taking the large coupling limit one used renormalization
group and symmetry arguments to determine the limiting theory.  These arguments seem
perfectly sensible {\it as long as it is true that the correct low energy degrees
of freedom have been completely identified}.  That is, the construction assumes
that the low energy degrees of freedom of the strongly coupled Higgs branch are just
the classical variables which parametrize that branch.

In ordinary quantum field theory, arguments like this can break down because of
the formation of bound states.  The true low energy degrees of freedom in a regime
not amenable to perturbation theory are not simple combinations of the underlying
degrees of freedom.  Rather, in the infrared limit, the description of these states as
composites of underlying degrees of freedom becomes singular, and they must be
introduced by hand as elementary fields.  One need look no further than the description
of the pion in the chiral limit of QCD for an example of this phenomenon.  In such
a context, nonrenormalization theorems about the Lagrangian of the underlying
fields may be misleading.   

What makes this particularly relevant in the present context is that we are trying
to describe highly composite states - the finite longitudinal fraction states of
a holographic theory.   It seems perfectly plausible that the bound state wave
functions become singular in the limit in which gravity decouples.  If that is the 
case, then the bound states may have to be introduced by hand as extra degrees of
freedom.  The existence of an effective theory which describes the interactions
of fivebranes in M theory in the limit that the Planck mass goes to infinity
does not by itself guarantee that this limit can be naively taken in the 
Lagrangian of the fundamental degrees of freedom.  This seems particularly 
worrisome in a holographic theory, in which {\it all} of the low energy
states are infinitely composite bound states of the fundamental degrees of
freedom.  

This might seem like so much nitpicking, but there is at least one context
in which we can see that the construction of \aharony\ fails.  For $k = 1$,
the $(0,2)$ field theory is the theory of a free tensor multiplet.  In this
case the instanton moduli space is singular and we cannot make sense of
its quantum mechanics, without providing further prescriptions about how
to deal with the singularities.  One could adduce this as evidence that the whole
Matrix model approach to uncompactified M theory in the presence of 
longitudinal fivebranes is wrong, but it seems more likely that the
failure has to do with the singularity of bound state wave functions
in the limit $\lp \rightarrow 0$.  After all, the matrix model contains
no other length scales besides the Planck scale.  It thus seems quite
reasonable that the matrix model contains bound states corresponding to
higher longitudinal momentum modes of the tensor multiplet for all
finite values of the Planck scale, but that the description of these states
as bound states of D0 branes becomes singular in the $\lp \rightarrow 0$ 
limit.  

The singular $k = 1$ case lies just beneath the surface even for $k > 1$.
The $U(k)$ $(0,2)$ theory has a moduli space.  At generic points on this
moduli space the theory contains several infrared free tensor multiplets.
Like all backgrounds in the IMF, this moduli space is described by 
a change in the Hamiltonian.  It corresponds to adding masses to the 
fundamental hypermultiplets.  Generic points in moduli space seem to be
infected by the $k=1$ disease.

Finally, we note that even the description of the origin of the $(0,2)$ 
Coulomb branch by $SU(k)$ instanton moduli space quantum mechanics may be 
singular.  In this case there are nonsingular instantons, but the
boundaries of moduli space corresponding to \lq\lq zero scale
size instantons \rq\rq are a potential source of singularity and
ambiguity in the quantum mechanics.  The attempt to use a closely related
moduli space to study the properties of H monopoles in heterotic string
theory \ref\jeff{J.P.Gauntlett, J.A.Harvey, hep-th/9407111,
9403072.} encountered ambiguities associated with zero scale
size instantons which could only be resolved by an appeal to an underlying
string theory.  

These remarks should not be considered a definitive critique of the beautiful
scenario of \aharony\ and \ed\ , but merely a cautionary statement which
suggests a direction for further study.

\subsec{Compactifications with Less Than Maximal SUSY}

When one begins to contemplate the breaking of SUSY in the matrix model, one idea which
immediately suggests itself is to change the base manifold of the SYM theory into one
with nontrivial holonomy, which preserves only part of the sixteen SUSY generators.
It is not immediately obvious that this is the right prescription, but in the case
of four compact dimensions, Berkooz and Rozali \ref\br{M.Berkooz,
M.Rozali, hep-th/9705175.} have provided arguments
that indeed the entire moduli space of M theory on K3 can be understood in terms of
the $(0,2)$ theory compactified on $K3 \times S^1$.  

We will not pursue this idea here, but instead discuss constructions of the heterotic
string following a line of thought motivated by the work of Horava and
Witten\ref\hw{P.Horava, E.Witten, Nucl.Phys. B460 (1996) 506.}.
We begin by studying an isolated Horava-Witten domain wall in eleven dimensional spacetime.
This produces a model written down by Daniellson and Ferretti
\ref\df{U.Danielsson, G.Ferretti, hep-th/9610082.} in the context
of Type I' string theory.  It was first introduced in Matrix Theoryby Kachru and
Silverstein\ref\ks{S.Kachru, E.Silverstein, Phys.Lett. B396 (1997) 70,
hep-th/9612162.}.  We will approach via a trick suggested by Motl
\ref\motl1{L.Motl, hep-th/9612198.}.
Namely, we \lq\lq mod out \rq\rq by the Horava-Witten symmetry of the original model
\eqn\modsymm{X^1 \rightarrow - (X^1 )^T}
\eqn\modsymmb{X^a \rightarrow (X^a )^T}
\eqn\modsymmc{\Theta \rightarrow \gamma^1 \Theta^T}
The transpositions in these formulae are the analog of the $A_3 \rightarrow - A_3$
operation on the SUGRA three form in the Horava-Witten transformation.  Indeed, the three
form couples to membrane world volumes, so a reversal of it is equivalent to orientation
reversal on the membrane world volume.  Recalling that the volume form on the spatial
membrane volume in light cone gauge is replaced by the commutator in Matrix Theory, we see
that transposition is the appropriate analog of three form reversal.

Modding out by this symmetry means restricting the variables so that the symmetry is 
equivalent to a gauge transformation and does not act on gauge invariant quantities.
There are two inequivalent restrictions, to either orthogonal or symplectic gauge groups.
The appropriate one for the present discussion is the orthogonal group $O(N)$ while
a symplectic reduction (accompanied by reflection of five of the coordinates) describes
$T^5/Z_2$ \ref\reykimetal{N.Kim, S.J.Rey, hep-th/9705132; A.Fayyazudin,
D.J.Smith, hep-th/9703208.} \ns\ .   The orthogonal reduction is most easily understood in Type
I' language \df\ , \ks\ : here one is studying zero branes near an orientifold 
and the $O(N)$ symmetry arises as usual from the images of the zero branes.

The end result is a matrix model with $O(N)$ gauge group.  $X^1$ is restricted to be
an antisymmetric matrix.  It therefore has no zero mode and $X^1 = 0$ represents the
position of the HW domain wall.  The other $X^i$ are symmetric.  $\Theta$ can be 
decomposed as $\Theta = \theta \oplus \lambda$, where $\theta$ and $\lambda$ are
eight component spinors corresponding to positive and negative eigenvalues of $\gamma^1$
respectively.  $\theta$ is a symmetric matrix, while $\lambda$ is antisymmetric.

Only half of the SUSY generators, those obeying $\gamma^1 Q = - Q$, are preserved by the
symmetry.  Under these $X^i$ and $\theta$ transform as a supermultiplet, while
$X^1$, $\lambda$ and the Lagrange multiplier which enforces the Gauss Law constraint
for $O(N)$ transform as a gauge multiplet.  This 
quantum mechanical SUSY is closely related to the $(0,8)$ SUSY of the heterotic
string.  Classical supersymmetric \lq\lq vacuum \rq\rq states of the system
have commuting $X$ coordinates.  In the Type I' picture these correspond to moving 
D0 branes away from the orientifold.  If the system is to have an interpretation as
a localized wall embedded in eleven dimensional spacetime, these must be zero energy 
states of the quantum mechanics.  More precisely, the commuting values are 
Born-Oppenheimer coordinates representing the slow motions of supergravitons
away from the wall.  The Born-Oppenheimer potential for these modes should vanish.

In fact, as first pointed out in \df\ , the potential does not vanish.  This is
somewhat surprising in view of the fact that the flat directions of the classical
potential are invariant under the $(0,8)$ SUSY.  The resolution of this paradox
was presented in \ref\bssilv{T.Banks, N.Seiberg, E.Silverstein, hep-th/9703052.}: the superalgebra only closes up to gauge transformations.
Classically, the flat directions are gauge invariant.  However, quantum corrections
change this situation.  The choice of a flat direction breaks $O(N)$ to a product
of $U(N_i )$.   For simplicity, consider $N$ even, and the breaking to $U(N/2)$.  This 
corresponds to moving all supergravitons away from the domain wall by the same amount.

 The fermionic modes which \lq\lq get mass \rq\rq from this breaking
are charged under the $U(1)$ subgroup which represents motion away from the
wall.  There are couplings of the form $\psi A_0 + X^1 \psi$, where $\psi$ is
real.  These are chemical potentials for the $U(1)$ charge.  Thus, the ground state
of these modes is charged, and a term $\vert A_0 + X^1 \vert$ is induced in the
 effective action.  This is a supersymmetric term which acts like the $0 + 1$ dimensional
analog of a Chern-Simons term.  It contains a linear potential for motion away from the
wall.   In the presence of such a term, the system does not contain a sector
describing the free propagation of supergravitons far away from the wall.

\df\ pointed out that the addition of $16$ real fermions transforming in the $N$
of $O(N)$ cancelled this term precisely.  In the Type I' picture, these represent
D8 branes (and their images) sitting on top of the orientifold, and the cancellation
can be viewed as the quantum mechanical version of the cancellation of the linearly
rising dilaton field of an orientifold found by Polchinski and Witten
\ref\polchwit{J.Polchinski, E.Witten, Nucl.Phys. B460 (1996) 525,
hep-th/9510169 }.
\ks\ showed that this system contained $E8$ gauge bosons propagating on the
domain wall, when one took into account both even and odd $N$.

To obtain the full Horava-Witten picture of the heterotic string and its relation to
an eleven dimensional theory compactified on $S^1 / Z^2$ we must apply the Horava-Witten projection to the $1 + 1$ dimensional $U(N)$ SYM theory which describes M
theory compactified on a circle.  We obtain a $(0,8)$ SUSY gauge theory with 
a left moving gauge multiplet and a right moving matter multiplet in the symmetric
tensor of $O(N)$.  This theory has an anomaly \ref\rk{N.Kim, S.J.Rey,
{\it op. cit.}} \bssilv\ .  For large $N$, the only
way to cancel it is to add $32$ real left moving fermions in the $N$ of $O(N)$.  
Once these are introduced, we must also choose the gauge bundle of $O(N)$ on the circle.
$O(N)$ is the subgroup of $U(N)$ which commutes with the HWM projection.  In its
action on the original fields of the matrix model, it is equivalent to $SO(N)$,
but in the fundamental representation the transformation which acts as $-1$ is 
no longer trivial.  Thus, we must choose boundary conditions for each of the $32$ 
fundamental fermions.  This choice is fixed uniquely by the requirement that 
in the limit in which the gauge theory becomes weakly coupled, and the circle which
it lives on is small, the system reduce to two copies of the SUSY quantum mechanics of
$\df$.   

In the indicated limit, the field $X^1$ becomes a classical variable.  It is the
covariant derivative for the constant gauge connection $A_1$ representing the Wilson
loop around the circle.  The eigenvalues of $R_1 A_1$ represent the positions of
supergravitons on the Horava-Witten interval.  The Hamiltonian for the
$32$ fermions is 
\eqn\fermham{\chi^A_i R_1 ({\partial \over i\partial\sigma} - A_1)_{ij} \chi^A_j}
where $A$ runs from $1$ to $32$.  As $R_1$ gets large, most of the modes of
these fermions go off to infinity.  The exceptions are when $A_1 = 0$ ($A_1 = \pi$)
where periodic (antiperiodic) fermions have zero modes.  Thus, in the large $R_1$ 
limit, we obtain a system with two domain walls at $0$ and $\pi R_1$ when half of the
fermions are chosen to have periodic and the other half antiperiodic boundary
conditions.  With this choice, subsystems far from the walls will not feel the
effects of the walls.

The above analysis was presented in \ref\bm{T.Banks, L.Motl, hep-th/9703218.} , and similar results were also
obtained by \ref\rletc{D.A.Lowe, hep-th/9702006, hep-th/9704041;
S.J.Rey, hep-th/9704158, P.Horava, hep-th/9705055.}.  The limit of weak string coupling (strong gauge coupling)
was also studied, and shown to correspond to the heterotic string.  The basic idea is
again that the strong coupling limit forces one on to the moduli space of 
commuting matrices.  This can be represented as the space of diagonal matrices
modulo permutations.  However the system also contains the $32$ massless fundamental
fermions (the continuous gauge group is completely broken on the moduli space, so these
fermions have no gauge interactions) .  There is a residual gauge symmetry, $Z_2^N$, 
of multiplication of each of the $N$ components of the fermions by $-1$.  The full
discrete gauge symmetry of the moduli space conformal field theory is the semidirect 
product of the permutation group $S_N$ with this group of reflections.  The
conjugacy classes are products of cyclic permutations and products of cyclic permutations 
and a reflection which changes the sign of the last element 
of each cycle in the fundamental representation.

As in the derivation of IIA string theory, states with energies of order $1/N$ are obtained
from sectors of the orbifold conformal field theory with cycle lengths of order $N$.
Multicomponent (diagonal matrix or vector) fields with such twisted boundary conditions
correspond to single component fields on an interval with length of order $N$.
We will present formulae here only for the $32$ fermion fields

Let $D^a_b$ be the $32\times 32$ matrix which multiplies the periodic
fermions by $1$ and the antiperiodic fermions by $- 1$.  We wish to find
vector valued fields which satisfy the boundary conditions
\eqn\cycbc{\Psi^a_i (\sigma + 2\pi ) = D^a_b Z_i^j \Psi^b_j (\sigma ).}
or
\eqn\cycbcb{\Psi^a_i (\sigma + 2\pi ) = D^a_b Z_i^j \delta_j^k
(-1)^{\delta_j^L} \Psi^b_k (\sigma ),}
where $L$ is the length of the cycle.
In keeping with the multistring Fock space interpretation of the general
boundary condition, we can, without loss of generality restrict
attention to cyclic permutations $Z$.  Note however that in doing so we
must also consider all possible values  $L \leq N$ for the length of
the cycle.  In particular, we must have cycles of both even and odd lengths.

The general solution of these boundary conditions in terms of single
component fields defined on a large circle , depends on the parity of
$L$.
We write
\eqn\cycsoln{\Psi^a_i (\sigma ) = (D^{(i - 1)})^a_b \psi^b (\sigma +
2\pi (i-1) )}
Then, the 32 component fermion $\psi^a$ has to satisfy
\eqn\cycsolnb{\psi^a (\sigma + 2\pi L) = (D^L)^a_b \epsilon
\psi^b(\sigma ).}
$\epsilon = \pm 1$ depending on whether we are in the sector \cycbc\ or
\cycbcb\ .  If $L$ is even, we obtain a sector with $32$ fermions, all
of which have periodic (P) or antiperiodic (A) 
boundary conditions, depending on
$\epsilon $.    If $L$ is odd, the fermions split into two groups with
either $AP$ or $PA$ boundary conditions, depending on $\epsilon$. 

We must also impose gauge invariant projectors on physical states.  The
two relevant conjugacy classes of gauge transformations are the cyclic
permutation, and the overall $O(1)^L$ transformation on the fermions.
The latter is equivalent to a reflection of all $32$ fermions $\psi^a$.
The former splits into two transformations in the large $L$ limit.  Part
of it becomes the infinitesimal translation operator on the \lq\lq long
\rq\rq heterotic strings.  However, as a consequence of the form of
Equation \cycsoln\ the cyclic permutation also acts by multiplying 
$\psi^b$ by the matrix $D$.  Thus, this gauge transformation also
includes a discrete operation on the heterotic fermions which multiplies
half of them by $-1$.    
Thus, we reproduce the \lq\lq internal \rq\rq GSO projection of the
heterotic string.

Compactification of the heterotic string on $T^d$, presents further complications.
\bm\ presented a prescription in which the resulting matrix model was a $U(N)$ gauge theory
on $S^1 \times T^d / Z_2$. The $32$ fermions are distributed among the orbifold
circles of this manifold\foot{Motl also suggested to the author that moving the
fermions around was equivalent to putting in $E_8 \times E_8$ Wilson lines.} .
This prescription can at best describe certain points in the moduli
space of compactified heterotic strings.  Attempts to describe the rest
of moduli space in this formalism are beset by problems of anomaly
cancellation. 
Horava\ref\petr{P.Horava, hep-th/9705055.}
suggested that they be cured by adding Chern-Simons terms to the bulk theory, but did
not present a supersymmetric Lagrangian accomplishing this task.  Kabat
and Rey \ref\kr{D.Kabat, S.J.Rey, hep-th/9707099.}
have constructed such a Lagrangian for the case of compactification on a single circle.

It is my present opinion that the best way to approach heterotic compactifications
is by using the duality relation to Type IIA strings on $K3$.  
Aspinwall\ref\psa{P.S.Aspinwall, hep-th/9707014.} 
has recently presented a persuasive argument based on string-string duality that
heterotic string theory is best viewed as a singular limit of $K3$ compactifications 
of the IIA string.  Berkooz and Rozali, \br\ (see also \ref\gov{
S.Govindarajan, hep-th/9705113, hep-th/9707164.}) presented a
Matrix Theory of compactification of M theory on $K3$, which contains various
heterotic compactifications as singular limits.  Many generic features of the
nonperturbative physics become obscure in the heterotic language.  Thus it seems
best to approach the heterotic theory as a member of a larger family of nonsingular
compactifications rather than to try to force all of moduli space into heterotic
language.

\subsec{Summary of Compactification}

The problem of compactification of Matrix Theory has turned out to be
fascinatingly complex.  Current results seem to indicate that the
theory is nonlocal and contains degrees of freedom which vary with the
compactification.  The eleven dimensional limiting theory does not
contain the full complement of canonical variables and we have not
yet found the maximally compactified theory (in the sense of the
theory with the full set of degrees of freedom).  The idea that different
compactifications might correspond to different ways of taking the
large $N$ limit of the original matrix quantum mechanics \bfss\ seems
much less plausible in view of the replacement of the SYM prescription
for compactification by more exotic theories on the four and five torus.

There are very clear indications that spacetime is not a fundamental
notion in the theory.  Rather, the SUSY algebra with its full complement
of central charges seems to be the object which makes sense in all regions
of moduli space\foot{The idea that the SUSY algebra is {\it central}
to M theory originates with P. Townsend \ref\ptown{P.K.Townsend,
hep-th/9507048 .} and has also been 
explored by Bars \ref\ibars{I.Bars, hep-th/9604139.}.}.  
Geometry emerges via a Born-Oppenheimer approximation in regimes where
certain BPS charges define a dense spectrum of low energy states.   
Compact geometries may be characterized solely by the spectrum of
wrapped BPS states to which they give rise.  That is, these are data
characterizing the geometry which have meaning even in those regimes
where classical geometric notions are not valid.  Two geometries with
identical BPS spectra may be the same (a generalized notion of mirror
symmetry) in Matrix Theory.

There is clearly some way to go before we have a full picture of
compactification.  The results of Berkooz and Rozali \br\ suggest that
a complete understanding of toroidal compactification with
maximal SUSY may go a long way towards pinning down the prescription
for compactification with partial SUSY breaking.  Matrix Theory on
K3 is essentially determined by Matrix Theory on a four torus.  

Of course, once we reach compactifications with only four real SUSY
charges, new issues will certainly arise.  In this case we do not
expect to have a moduli space of vacua, and issues of cosmology
and the cosmological constant will arise.  It is likely that we will
not be able to study this regime without freeing ourselves from
the light cone gauge.

There are indications that certain cosmological issues may have to be
dealt with even in cases of more SUSY.  Even if the current difficulties
of compactification on a six torus are resolved, more problems await us
on the $7,8$ and $9$ tori.  On the seven torus, long range forces
between individual D0 branes grow logarithmically, and things become
more serious as we go down in the number of noncompact dimensions. 
Finally, on the $9$ torus, we are faced with an anomalous SYM theory.
Although we have every reason to believe that $9+1$ dimensional SYM
theory is not the full description of $T^9$ compactified M theory, 
it should be a valid description in the regime where all the radii are
larger than the Planck scale.  Anomalies in low energy effective
theories have historically signified true problems with the dynamics.
Susskind and the present author have speculated that this anomaly is
related to the behavior of $T^8$ compactified string theory
which they observed in \banksuss\ .  There, it was shown, by examining
the classical low energy field equations, that the string theory had no
sensible physical excitations of the compactified vacuum.  In a first
quantized light cone description, like Matrix Theory, we should find no
states at all.  This is precisely the message of the anomaly.  It
implies a Schwinger term in the commutator of gauge generators which
precludes the existence of solutions of the physical state condition.

The authors of \banksuss\ suggested a cosmological interpretation of
their results. In a completely compactified theory one cannot ignore
quantum fluctuations of the moduli.  Furthermore, since
\ref\moorehorne{J.H.Horne, G.Moore, Nucl. Phys. B432, (1994), 109,
hep-th/9403058. } moduli space is of finite volume,
the {\it a priori} probability of finding any sort of large volume
spacetime is negligible.  The system quantum mechanically explores its
moduli space until some fluctuation produces a situation in which a
classical process (inflation?) causes some large spacetime dimensions to
appear.  It then rolls down to a stable equilibrium point.  The absence
of physical excitations in the toroidally compactified vacuum make it an
unlikely (impossible?) candidate for this quiescent final state.  
Thus, the failure to find a satisfactory nonperturbative formulation of
M theory with complete toroidal compactification may help to resolve
one of the primary phenomenological puzzles of string theory: why we do
not live in a stable vacuum with extended SUSY.  

\newsec{Discrete Light Cone Quantization}

One of the remarkable features of the results described in the previous section
is that we obtained most of them without taking the large $N$ limit.  
In particular, U duality was a property of the finite $N$ theory.  {\it A priori}
there is no reason for this to be so.  Our arguments that the matrix model was all
of M theory were valid only in the large $N$ limit.  

Susskind \ref\dlcq{L.Susskind, hep-th/9704080.} has provided a conjectural understanding of the remarkable
properties of the finite N matrix models by suggesting that they may be the Discrete
Light Cone Quantization (DLCQ)\ref\pauli{T.Maskawa, K.Yamawaki,
Prog. Theor. Phys. 56 (1976), 270. For a review see S.J.Brodsky,
H.-C.Pauli, in {\it Recent Aspects of Quantum Fields}, ed. H.Mitter,
H.Gausterer, {\it Lecture Notes in Physics} (Springer-Verlag, Berlin,
1991) Vol. 396.} of M theory.  In quantum field theory
one often replaces IMF quantization by light cone quantization.  Rather than taking
an infinite boost limit of quantization on a spacelike surface, one quantizes
directly on a light front.  This procedures shares the simplifications of IMF
physics that result from positivity of the longitudinal momentum, but does not
require one to take a limit.  Within the framework of light cone quantization, one
can imagine compactifying the longitudinal direction.  The theory breaks up into 
sectors characterized by positive integer values $N$ of the longitudinal momentum.
The idea of DLCQ is that the sectors with low values of $N$ have very simple
structure.  In field theory, the parton kinetic energies are very simple and
explicit, and the complications of the theory reside in interactions whereby
partons split into other partons of lower longitudinal momentum.  In the sector with
$N = 1$, this cannot happen, so this sector is free and soluble.  In sectors with small
values of $N$ the number of possible splittings is small, and in simple field theories
the Hamiltonian can be reduced to a finite matrix or quantum mechanics of a small
number of particles.  Note that these simplifications occur despite the fact that
we keep the full Hamiltonian and make no approximation to the dynamics.  As a 
consequence, {\it any symmetries of the theory which commute with the longitudinal
momentum are preserved in DLCQ for any finite N}.  This is the basis for Susskind's
claim.   It is manifestly correct in the weakly coupled IIA string limit of M theory
(at least to the order checked by \dvv\ ).  Since we have no other nonperturbative
definition of M theory to check with,  Susskind's conjecture cannot be checked
in any exact manner. However, Seiberg \natiproof\ has recently given a
formal argument that Matrix Theory is indeed the exact DLCQ of M theory.

One of the most interesting areas of application of the DLCQ ideas is the matrix
description of curved space.  Although I do not have space to do justice to this
subject here, I want to make a few comments to delineate the issues.

An extremely important point is that spatial curvature always breaks some SUSY,
so that many things which are completely determined by maximal SUSY are no longer
determined in curved space.  Consequently, the Matrix Theory Lagrangian in
a curved background cannot be written down on the basis of symmetries alone.
Related to this is the fact that for a sufficiently small residual SUSY algebra,
supersymmetry alone does not restrict the background to satisfy the equations of motion.
If we succeed in constructing Matrix Theory on curved backgrounds, what will
tell us that the background must satisfy the equations of motion?

One answer to this question can be gleaned from the nature of the finite $N$ theory
in perturbative string theory.  Finite $N$ can be thought of as a kind of 
world sheet cutoff.  In this way of thinking about things, the matrix field
theory background does not have to satisfy the equations of motion.  Rather, matrix
field theory backgrounds will fall into universality classes.  Requiring longitudinal
boost invariance in the large $N$ limit will determine that the effective large $N$
background will satisfy the equations of motion.  It is only in this limit that
the correct physics will be obtained.
Thus, Fischler and Rajaraman\ref\willy{W.Fischler, A.Rajaraman,
hep-th/9704123. }
argue that the difficulties uncovered by Douglas, Ooguri and Shenker
\ref\dos{M.R.Douglas, H.Ooguri, S.Shenker, Phys.Lett. B402 (1997) 36,
hep-th/9702203. }
in the description of Matrix Theory on an ALE space by quantum mechanics
with eight SUSYs and a Fayet-Iliopoulos term, will disappear in the large $N$ 
limit.

I believe that this point of view is correct, but DLCQ suggests a complementary
strategy.  Namely, among all of the members of a \lq\lq large $N$ universality
class \rq\rq there should be one (which corresponds to the DLCQ of the exact theory)
in which all physics unrelated to longitudinal
boosts or full Lorentz invariance is captured correctly at finite $N$.
In particular, one might argue (but see the discussion below) that 
at distances larger than $\lp$, {\it transverse} geometry should be
that determined by SUGRA even in the finite $N$ theory.
Douglas\ref\mrd{M.R.Douglas, hep-th/9703056, 9707228; M.R.Douglas,
A.Kato, H.Ooguri,\break hep-th/9708012.} has
suggested a strategy for discovering the correct DLCQ Lagrangian for finite $N$ matrix
theory in Kahler geometries,  and he and his collaborators have begun to explore
the consequences of his axioms for matrix geometry.  The most important of these
axioms is that large distance scattering amplitudes determined by the matrix model
should depend on the correct geodesic distance in the underlying manifold.  In
a beautiful recent paper, \ref\do?{M.R.Douglas, A.Kato, H.Ooguri, {\it
op. cit.}} Douglas {\it et. al.} have shown that
the Ricci flatness conditions follow from Douglas' axioms.  This is a very promising
area of research and I expect more results along these lines in the near future.

Another puzzle for the DLCQ philosopy is provided by compactification of
Matrix Theory on a two torus.  We showed above how Type IIB string
theory arises from the matrix model.  However, if we add a real part to
the complex structure parameter $\tau$ of the torus an interesting
paradox arises.  Along the moduli space, Type IIB perturbation theory is
obtained by writing a Kaluza-Klein expansion of the $2+1$ dimensional
matrix fields as an infinite set of $1+1$ dimensional fields living on
the long cycle of the SYM torus.  Most of these $1+1$ dimensional fields
have masses which depend on $Re\ \tau$.  The masses are of order
$g_S^{-1}$, and although the $Re\ \tau$ dependence is of subleading
order, standard notions of effective field theory lead one to expect
$Re\ \tau$ dependence in finite orders of the $g_S$ expansion.  We do
not know enough about the superconformal field theory which underlies
the finite coupling IIB string theory to prove that this is so, but we
certainly have no proof to the contrary.  Thus, it appears likely that
the finite $N$ matrix model does not give the DLCQ of the IIB
perturbation series.

Some examples from field theory may shed light on this puzzle.  Indeed,
in quantum field theory there would appear to be a number of
inequivalent definitions of the DLCQ of a given theory.  We can for
example examine the exact Hilbert space of the theory quantized on a
lightlike circle and restrict attention to the subspace with
longitudinal momentum $N$.  On the other hand, we can choose a specific
set of canonical coordinates and restrict attention to states in the
Fock space defined by those coordinates which have momentum $N$.  The
example of $SU(K)$ QCD with ({\it e.g.}) $K > N$ shows that these two
restricted spaces are not equivalent.  Baryons with longitudinal
momentum $\leq N$ are included in the first definition, but not in the
second.  It is also evident that the Hilbert space defined by DLCQ of a
canonical Fock space is not invariant under nonlinear canonical
transformations.  Perhaps this can explain the paradox about DLCQ of IIB
theory described in the previous paragraph.

On the other hand, DLCQ does seem to preserve the duality between IIA
and IIB string theory as an exact duality transformation of $2+1$
dimensional SYM theory, so perhaps the intuitions from field theory are
not a good guide.

A number of calculations of scattering amplitudes in situations of lower
SUSY are now available in DLCQ\ref\dosdr{M.Douglas, H.Ooguri,
S.H.Shenker, Phys. Lett. B402, (1997), 36, hep-th/970203; M.Dine, A.Rajaraman, hep-th/9710174.}.  They disagree with the
predictions of tree level SUGRA.  In view of Seiberg's derivation of the
Matrix Theory rules, this appears to pose a paradox.
 Douglas and Ooguri\ref\do{M.R.Douglas, H.Ooguri, hep-th/9710178.} have
recently analyzed this situation and described two possible ways out of
this paradox.  The first, which is the one favored by these authors, is
that our extraction of the Hamiltonian for the DLCQ degrees of freedom
from weakly coupled Type IIA string theory, is too naive.
Renormalizations due to backward going particles will renormalize the
Hamiltonian in order to enforce agreement with SUGRA.

The other possibility is that the low energy limit of DLCQ M theory is
not tree level DLCQ SUGRA.  Indeed, the derivation of tree level SUGRA
as the low energy limit of uncompactified M theory, uses eleven
dimensional Poincare invariance in a crucial way.  The very existence of
two different, unitary , super Galilean invariant amplitudes (as
apparently shown by the calculations of \dosdr\ ) proves that the light
cone symmetries are insufficient to obtain this result after DLCQ.  
It is not clear what, if any, consequences follow for finite $N$ DLCQ
from the requirement that the S matrix be the DLCQ of a fully Lorentz
invariant theory.  Thus, I do not see a proof of the equivalence of the
two systems at low energy and finite $N$.\foot{One should mention a
third possible resolution of these paradoxes, namely that the large $N$
limit of DLCQ M theory does not converge to uncompactified M theory at all.}

Furthermore, Seiberg's results about the six torus compactification are
close to being a counterexample to the proof that the two low energy
limits are the same.  Based on rather general BPS arguments, Seiberg
proves that for finite $N$, M theory on a six torus contains objects
with a continuous spectrum starting at zero which do not have a
conventional interpretation in terms of the M theory spacetime.
Further, he argues that these states do not decouple from the 
states which carry momentum in the M theory space time (we have
described these arguments above).  As a consequence, the low energy
limit of the DLCQ M theory S-matrix contains processes in which this non
spacetime continuum is excited.  Clearly there is no analogous process 
in tree level SUGRA.  

In order to avoid this obvious inequivalence between the two systems one would
have to argue that the corrections to the naive Matrix Theory dynamics
removed the coupling between the two kinds of degrees of freedom \foot{One
cannot remove the continuum spectrum.  It consists of wrapped BPS six
branes. I have not found an argument based on SUSY alone which
guarantees that the low energy coupling is that described by Seiberg,
but since it follows from minimal SUGRA coupling in the T dual picture
it seems difficult to avoid.}.  This seems unlikely.  
I believe that the simplest conclusion from this example is that after
DLCQ, SUGRA and M theory simply do not have the same low energy limit.

This greatly restricts the {\it a priori} tests of Matrix Theory which
one might imagine doing for finite $N$.  The finite $N$ theory will have
the duality and SUSY properties we expect of the full theory, but it now
seems unlikely that it will reproduce much of the correct physics for
small $N$.  I remind the reader that many of the properties of the
Matrix Theory which we have exhibited, the existence of membranes and of
the full Fock spaces of supergravitons and Type II strings for example,
depended crucially on taking the large $N$ limit.  It seems to me that 
one of the most
important hurdles to be overcome in the development of the theory, is
learning how to take this limit in an elegant and controlled manner.
In an ideal world one would hope to be able to formulate the theory
directly at infinite $N$. Clearly there is much to be understood in this area. 

\newsec{BPS Branes as Solitons of the Matrix Model  }

The material in this
section is a brief summary of \bss\ , \bd\ .  I am including it
mostly in order to provide a more up to date understanding of the material in
these papers and the reader should consult the original papers for details.
Much of the progress in string duality has come from an understanding of the
various BPS p-branes that string/M theory contains.  Branes with $0 < p \leq 5$ 
can be understood as incarnations of the M theory 5 brane or two brane.  
Of the rest, the Horava-Witten end of the world ninebrane and the M theory 
Kaluza-Klein monopole (the D6 brane of IIA string theory) play significant roles.
The former has been described in our discussion of the heterotic string.
The latter will clearly be a key player in the description of Matrix
Theory on $T^6$,
which is as yet poorly understood.
The D7 brane made a brief appearance in the origin of F theory.  
We will concentrate here on the fivebrane and membrane of M theory.  

We have described how finite uncharged membranes appear in the eleven dimensional
matrix model.  Wrapped membranes played a role in our discussion of the normalization of
the parameters of the SYM theory compactified on a torus.  They are configurations of
nonzero magnetic flux in the SYM theory.  In four toroidal dimensions, where the SYM
theory is replaced by the $(2,0)$ field theory, they are configurations of
electric (which is the same as magnetic because of self duality) flux of 
the two form gauge field.  The wrapped membrane charges correspond to components of
the two form electric flux in four out of the five dimensions of the five torus on
which the $(2,0)$ theory lives.  The components involving the fifth toroidal direction 
and one of the other four represent Kaluza-Klein momenta in the M theory spacetime.
This description is obviously not invariant under the $SL(5,Z)$ duality symmetry of
the $(2,0)$ theory.  It is valid only in the region that the fifth toroidal direction 
is much smaller than the other four.  It then defines the eleven dimensional Planck scale, \brs\
.  In a generic region of the space of backgrounds,
we simply have BPS charges in the $10$ dimensional second rank antisymmetric tensor
representation of $SL(5,Z)$.  The breakup into $6$ wrapped membrane charges and $4$ Kaluza-Klein
momenta is only sensible in regions where an eleven dimensional M theoretic spacetime
picture becomes valid.  In any such regime, the $(2,0)$ theory is well approximated by SYM.
At a fundamental level, the wrapped transverse membrane charges and the transverse momenta
are all part of the BPS central charge which appears in the anticommutator of a
dynamical and a kinematical SUSY generator.

In the limit of noncompact eleven dimensional spacetime, most of the SYM degrees of freedom
decouple, and the system becomes the super quantum mechanics which
describes M theory in eleven noncompact dimensions.
However, in the presence of one or more wrapped membranes we must keep enough of the SYM
degrees of freedom to implement the relation 
$Tr [X^m , X^n] = W_{mn}$, with $W_{mn}$ the membrane wrapping number.  This reproduces the
ansatz of \bfss\ .  As shown in \bss\ and elaborated upon in
\ref\finn{E.Keski-Vakkuri, P.Kraus, hep-th/9706196.}\ one can also study
low energy fluctuations around these configurations.  The enhanced gauge symmetry which
obtains when two membranes approach each other, originally derived in the D-brane formalism
can be rederived directly from the matrix model.  This serves as the starting point
for the calculation of \pp\ .

One can also discuss membranes with one direction wrapped on the transverse torus and
the other around the longitudinal axis.  These are configurations which carry a BPS
charge which appears in the anticommutator of two dynamical SUSY generators, and carries
one transverse vector index.  As explained in \bs\ this charge is just the momentum of
the SYM field theory on the dual torus.  Indeed, the dynamical SUSY generators in the IMF
are, in those dimensions where the SYM prescription is the whole story, just the   
SUSY generator of the SYM theory in temporal gauge.  They close on the SYM momentum, up
to a gauge transformation (sometimes there are other central charges for topologically
nontrivial configurations).  

If the situation for membranes is eminently satisfactory, the situation for fivebranes is
more obscure.  Longitudinal fivebranes were first discussed in \bd\ .  These authors
observed that the D4 brane was the longitudinally wrapped fivebrane of M theory and
they could boost it into the IMF by considering its interactions with an infinite number of 
D0 branes.  This leads to SUSY quantum mechanics with eight SUSYs containing fields
in the vector multiplet and hypermultiplets in the adjoint and k fundamental
representations (for k fivebranes).   We have discussed this model above in the \lq\lq
matrices for matrices \rqq ansatz for the $(2,0)$ field theory.  As we will see this is apt
to be the proper definition of longitudinal fivebranes in eleven dimensional spacetime.

\grt\ and \bss\ tried to construct the longitudinal fivebrane as a classical solution of
the matrix model.  In particular, \bss\ observed that an object with nonzero values of
$Tr X^i X^j X^k X^l \epsilon_{ijkl}$ (with $\epsilon$ the volume form of some four dimensional
transverse subspace) would be a BPS state with the right properties to be the longitudinal
fivebrane.  Indeed, the BPS condition is $[X^i, X^j] = \epsilon_{ijkl} [X^k , X^l]$, 
which can be realized by the covariant derivative in a self dual four dimensional gauge
connection.  Thus \grt\ and \bss\ suggested that the limit of such a configuration in
the $4+1$ dimensional gauge theory, would be the longitudinal fivebrane.  Unfortunately,
this definition is somewhat singular for the case of minimal instanton charge (on a torus, this
gives an instanton of zero scale size).

  Perhaps it could be improved by going to the
$(2,0)$ theory and defining an instanton as a minimum energy state with the
lowest value of momentum around the fifth toroidal direction (which defines the Planck
scale).   At any rate, it is clear that in searching for infinite BPS branes in
eleven dimensions, we are discussing a limiting situation in which many degrees of freedom
are being decoupled.  This suggests that the best description will be the effective quantum 
mechanics of \bd\ which keeps just those degrees of freedom necessary to define the
longitudinal fivebrane.  

There is no comparable description of the purely transverse fivebrane.  Seiberg has 
explained why there is no SYM configuration which respresents it even in a singular
way.  Matrix Theory on a five torus is the theory of Type II NS fivebranes
in the limit of zero string coupling.  This theory {\it does} have a limit in which it
becomes $5+1$ dimensional SYM theory.  However, the wrapped M theory fivebrane
is a state of this theory which is translation invariant on the five torus and has
an energy density of order the cutoff scale.  Since it is not localized on the torus, it
does not give rise to a long range SYM field.  Nonetheless, by carefully taking the infinite
radius limit of such a wrapped fivebrane configuration we should find a description
in terms of the matrix quantum mechanics coupled to some other degrees of freedom, in the
spirit of \bd\ .  We do not yet understand enough about the theory of NS fivebranes
proposed in \ns\ to derive this construction.  One should also understand the
connection of these ideas to the proposal of \grt\ for constructing the transverse
fivebrane wrapped on the three torus.

\newsec{\bf Conclusions}

Matrix Theory is in its infancy.  It seems to me that we have taken some correct first
steps towards a nonperturbative formulation of the Hamiltonian which lies behind
the various string perturbation expansions.  It is as yet unclear how far we are from
the final formulation of the theory.   I would like here to suggest a plan for the
route ahead.  Like all such roadmaps of the unknown it is likely to lead to quite a few
dead ends and perhaps even a snakepit or two.  But it's the best I can do at the moment
to help you on your way if you want to participate in this journey.

First, we must complete the compactification of the maximally supersymmetric
version of the theory, and this for two reasons.  The present situation
seems to indicate new phenomena when there are six or more toroidally compactified 
dimensions.  Surely we will have to understand these in the controlled setting of
maximal SUSY if we are to understand them in more complicated situations.
In addition, the work of Berkooz and Rozali \br\ and Seiberg \ns\ suggests that  
at least the passage to half as many SUSYs is relatively easy once the 
theory has been formulated with maximal SUSY in a given dimension.
The crucial questions to be answered are whether the present impasse represents merely the
failure of a certain methodology (deriving the Matrix Theory Hamiltonian as a limit
of M theory in which gravity decouples), or signifies profoundly different physics
with more compactified dimensions.  As an extreme, one might even speculate that
compactification to a static spacetime with more than five (or six) compact dimensions
and maximal SUSY does not lead to a consistent theory.  
The clues that I believe are most important for the elucidation of this question
are all there in the SUSY algebra.  The lesson so far has been that the underlying
theory has degrees of freedom labelled by the \lqq finite longitudinal BPS charges \rqq
on the torus of given dimension.   The key here is the word finite.  When there are
seven or more compact dimensions then there are no finite BPS charges, as a consequence
of the logarithmic behavior of long range scalar fields in spacetime.  

Once we have constructed the maximally compactified, maximally supersymmetric Matrix Theory
we will have to understand systems with less SUSY.  As I have indicated, I suspect that
the first steps of this part of the program may be relatively straightforward.  One issue
which will certainly arise is the absence of a unique SUSY lagrangian for fewer than
$16$ SUSY generators.  Again I expect that this is probably irrelevant in the large $N$
limit, but that construction of the correct DLCQ Lagrangian for finite $N$ will require
new principles.  Perhaps duality will be sufficient.  The program initiated by Douglas
{\it et. al.} of studying DLCQ in noncompact curved space will undoubtedly teach us
something about this issue.

The conceptual discontinuity in this subject is likely to appear when we try to construct
systems with minimal four dimensional SUSY.  Low energy field theory arguments lead us to
expect that such systems have no real space of vacua.  We expect a nonvanishing superpotential
at generic points of the classical space of vacua which vanishes only in certain 
extreme limits corresponding to vanishing string coupling or restoration of higher SUSYs.  
Here is where all of the questions of vacuum selection (and the cosmological constant)
which we ask in weakly coupled string theory must be resolved.  Here also I expect the
disparity between the IMF and DLCQ points of view to be sharpest.  For finite $N$ we will
probably be able to construct a Matrix Theory corresponding to any supersymmetric 
background configuration, whether or not it satisfies the classical or quantum equations
of motion of string theory.  From the IMF point of view, these restrictions will arise 
from requiring the existence of the large $N$ limit, a generalization of the vanishing
$\beta$ function condition of weakly coupled string theory.  In particular, 
vacuum selection will only occur in the large $N$ limit.  On the other hand, if we succeed in
formulating principles which enable us to construct {\it a priori} the DLCQ Lagrangian, we
will find that these principles also pick out a unique vacuum state.  

There is however another feature of realistic dynamics which we will undoubtedly have
to cope with.  Astronomical evidence tells us that the world is not a static, time independent
vacuum state.  If Matrix Theory is a theory of the real world we should find that the
theory forces this conclusion on us: the only acceptable (in some as yet unspecified sense)
backgrounds must be cosmological\foot{As far as I know, the only hint of a reason for
theoretical necessity of a time dependent cosmology 
in string theory is a speculation which appeared in \banksuss\
.  I know the authors of that paper too well to give this speculation much credence.}.  
To the best of my knowledge, it has not been possible to formulate
cosmology in the IMF.   Thus, in order to deal with the real world we will have to reformulate
the basic postulates of Matrix Theory- or in other words \lqq to find a covariant
formalism \rqq .  

I have put the last phrase in quotes because I have emphasized that spacetime is a derived
rather than a fundamental quantity in the theory.  As a consequence, the notion of 
covariance remains ill defined at a fundamental level.  Perhaps, by adding enough
gauge degrees of freedom, we will find a formulation of the theory in which all
the various versions of spacetime appear together.  Then we might be able to formulate 
covariance in terms of a large symmetry group containing the diffeomorphism
groups of all versions of spacetime.  I believe that instead we will find a new notion which
replaces geometry and that covariance will arise automatically as a limit of the
invariance group of this new construct.  The only clue we have to this new notion of geometry
is the SUSY algebra.  The BPS charges of branes wrapped around cycles of a geometry
appear to be exact concepts in the theory we are trying to construct.  Is it possible that
the spectra of these charges is the entire content of the exact definition of geometry?
I have put a lot of thought into the discovery of a covariant formulation of the matrix
model, and have thus far come up with nothing.  I can only hope that this review will motivate
someone smarter than I to look at the problem.

Another issue connected to cosmology is the nature of time in Matrix
Theory and the
resolution of the famous Problem of Time in Quantum Gravity.  
It is commonplace in discussions of Quantum Gravity to point out that
conventional notions of time and unitary evolution must break down,
as a consequence of the very nature of a generally covariant integral
over geometries.  Yet at least on tori of dimension $\leq 5$ we seem to
have given a nonperturbative definition of a quantum theory with a
unique definition of time, and unitary evolution, which
reduces to general relativity at low energies.  For a space with some
number of noncompact, asymptotically flat,
 dimensions, the holographic principle provides us
with a convenient explanation of why this is possible.  In an
asymptotically flat spacetime, we can always introduce a unique (up to
Lorentz transformation) time at infinity.  The holographic principle
assures us that we can choose the hyperplane on which we project the
degrees of freedom of the theory, to lie in the asymptotically flat
region. Thus, it is perhaps not surprising that we have found a unitary
quantum theory for these cases.  But what of a completely compactified
cosmology? Is such a thing impossible in M theory?  The problems which
we encounter in trying to compactify the theory on high dimensional tori
suggest that there is something deep and puzzling going on, but we have
not yet been able to put our fingers on precisely what it is.

I suspect that the answers to these questions will only come when we
have found the beautiful mathematical structure which reduces to the
Riemann-Einstein geometry of spacetime in the low energy limit.  The
present formulation of Matrix Theory gives only tantalizing hints of
what that might be.  There has been much progress made in the last year,
but we can at best hope that the ultimate structure of the theory is near
our grasp.  Thirty years of string theory have taught us that new
puzzles constantly spring up to replace those which have just been
solved.  Only incurable optimists would dare to hope that the end is
in sight.  I continue to count myself among those fond and foolish
dreamers.

\listrefs
\end